\documentclass[10pt]{article}
\usepackage{amsmath}
\usepackage{amstext}
\usepackage{amssymb}
\usepackage{graphicx}
\def\ket#1{{|#1\rangle}}
\def\nn{\nonumber}
\title{Space time and the passage of time}
\author{George F R Ellis, University of Cape Town}
\begin{document}
\maketitle
\begin{abstract}
This paper examines the various arguments that have been put forward
suggesting either that time does not exist, or
 that it exists but it's flow is not real. I argue that (i) time both
 exists and flows; (ii) an Evolving Block Universe (`EBU') model of
 spacetime adequately captures this feature, emphasizing the key differences
 between the past, present, and future; (iii) the associated
 surfaces of constant time are uniquely geometrically and physically
 determined in any realistic spacetime model based in General
 Relativity Theory; (iv) such a model is needed in order to
 capture the essential aspects of what is happening in circumstances where initial
 data does not uniquely determine the evolution of spacetime structure because
 quantum uncertainty plays a key role in
 that development. Assuming that the functioning of the mind is based
 in the physical brain, evidence from the way that the mind apprehends the flow of time
 prefers this evolving time model over those
 where there is no flow of time.
\end{abstract}

\section{Space time and the Block Universe}\label{sec:BU}
In this section I briefly summarize the usual representation in
relativity theory of space-time as an unchanging block universe, and
the associated view that the change of time is an illusion.\\

The nature of spacetime in both special and general relativity has
lead some to a view that the passage of time is an illusion
\cite{SciAm12,But10,fqxi,fqxi_meet}. Given data at an arbitrary
time, it is claimed that everything occurring at any later or
earlier time can be uniquely determined from that data, evolved
according to deterministic local physical laws (this is formalized
in standard existence and uniqueness theorems \cite{HawEll73}).
Consequently, nothing can be special about any particular moment;
there is no special ``now'' which can be called the present. Past,
present and future are equal to each other, for there is no surface
which can uniquely be called the present.

Such a view can be formalized in the idea of a \emph{Block Universe}
\cite{Mel98,Sav01,Dav12}:\index{BU} space and time are represented
as merged into an unchanging spacetime entity, with no particular
space sections identified as the present and no evolution of
spacetime taking place. The universe just \emph{is}: a fixed
spacetime block, representing all events that have happened and that
will happen. This representation implicitly embodies the idea that
time is an illusion: time does not ``roll on" in this picture. All
past and future times are equally present, and the present ``now" is
just one of an infinite number.  Price \cite{Pri96} and Barbour
\cite{Bar99} in particular advocate such a position. Underlying
this,
 as emphasized by Barbour, is the idea that time-reversible Hamiltonian dynamics
 provides the foundation
 for physical theory in general and gravitation in particular.
Occasionally cosmology or astrophysics takes into account
time-irreversible physics, for example nucleosynthesis in the early
universe or the late phases of gravitational collapse, but the
notion of the present as a special time remains absent.

The problem with this view is that it is profound contradiction with
our experiences in everyday life, and in particular with the way
science is carried out. Scientific theories are developed and then
tested by an ongoing process that rolls out in time: initially the
theory does not exist; it is developed, tested, refined, finally
perhaps accepted: is it really plausible that all of this process is
an illusion, as some claim? Can it really be that ``Time is real but
flow is not''(Davies \cite{Dav12}), or ``Time does not exist''
(Barbour \cite{Bar99}, Rovelli \cite{Rov08})? If time is an
illusion, how can the mind generate this illusion, when (assuming
the validity of present day neuroscience) the mind is based in the
brain - a physical entity, governed by the laws of physics?

By contrast to this view, Broad already in 1923 \cite{Bro23} argued
that the true nature of spacetime is best represented as an
\emph{Emergent Block Universe} (EBU), \index{EBU} a spacetime which
grows and incorporates ever more events, ``concretizing" as time
evolves along each world line \cite{Ell06}. Unlike the standard
block universe, it adequately represents the differences between the
past, present, and future, and depicts the change from the
potentialities of the future to the determinate nature of the past.
This is the view I present in this chapter --- the claim ``time is
an illusion'' results from using an inadequate model of spacetime.

\section{Time and the Emerging Block Universe}\label{sec:EBU}
In this section I summarize the alternative representation of
space-time as an ever-growing emergent block universe, embodying the
view that the ongoing flow of time is a key physical aspect of
reality;
and relate this to the classical physics concept of the nature of time.\\

How do we envisage spacetime and the objects in it as time unrolls?
To motivate the EBU model of reality, consider the following
scenario \cite{Ell06}: A massive object has rocket engines attached
at each end that allow it to move either left or right. The engines
are fired alternately by a computer, which  produces firing
intervals and burn times based on a sensor activated by the random
decays of a radioactive element \cite{rand}. These signals select the actual
spacetime path of the object from the set of all possible paths. Due
to the quantum uncertainty inherent in radioactive decay
\cite{FeyLeiSan65,AhaRoh05,GreZaj06}, the realized path is not
determined by the initial data at any previous time: which potential
path becomes actual cannot be predicted, it is
determined as it happens (see Figure 1).\\

\begin{figure}[htb] \vbox{\hfil\scalebox{.5}
{\includegraphics{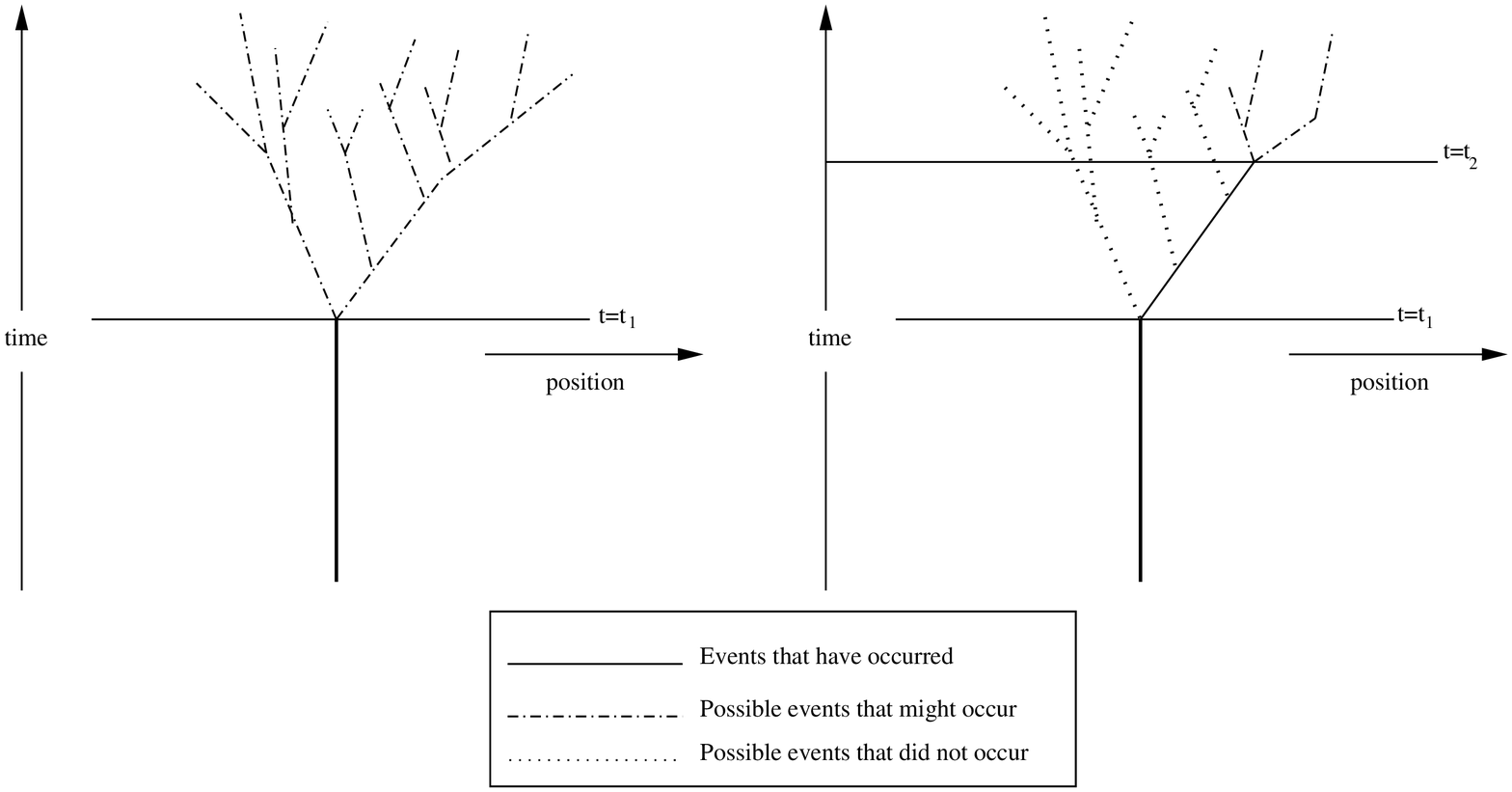}}\hfil} {\caption{\small{\emph{Motion
of a particle world line controlled in a random way, so that what
happens is determined as it happens. On the left events are
determined till time $t_1$ but not thereafter; on the right, events
are determined till time $t_2 > t_1$, but not thereafter. Spacetime
is unknown and unpredictable even in principle before it is
determined (because that choice is based on the randomness of
quantum decay of radioactive particles). The time at which it is
determined inexorably moves on, and given this physical context,
this unfolding cannot be stopped, changed, or reversed}.
 \label{fig1}}}}
\end{figure}

Because the objects are massive and hence produce spacetime
curvature,  spacetime structure itself is undetermined until the
object's motion is determined by the specific radioactive decay that
takes place. Instant by instant, spacetime structure changes from
indeterminate to definite. Thus a definite spacetime structure comes
into being as time evolves. The random element introduced through
the irreducible uncertainty of quantum events ensures that there is
no way the future spacetime can be predicted from the past: what
will actually happen is not determined until it happens.   Second by
second, one specific evolutionary history out of all  possibilities
is chosen, takes place, and becomes cast in stone (sometimes
literally). The future is uncertain and indeterminate until local
determinations have taken place at the spacetime event ``here and
now," designating the present on a world line at a specific instant;
thereafter this event is in the past, having become fixed and
immutable, with a continually new event on the world line
designating the present.

\begin{figure}[htb] \vbox{\hfil\scalebox{.5}
{\includegraphics{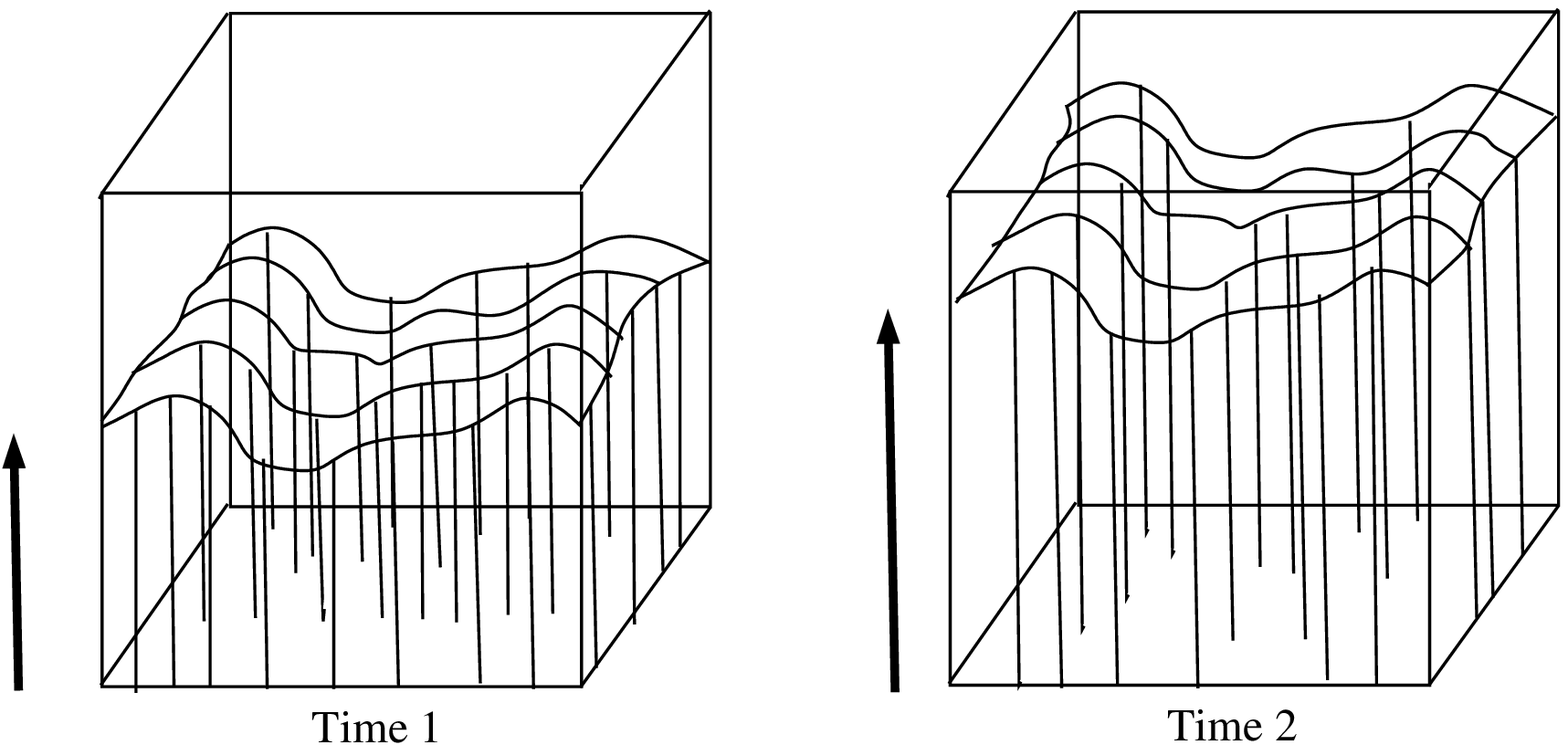}}\hfil} {\caption{\small{\emph{An
evolving curved space-time picture that takes the flow of time
seriously. Time evolves along each world line, extending the
determinate spacetime as it does so. You cannot locally predict
uniquely to the future from data on any constant `time' surface,
because of quantum uncertainty. This is true both for physics, and
for the spacetime itself: the developing nature of spacetime is
determined by the evolution of the matter in it. A key example is
the process whereby quantum fluctuations determine seed spacetime
inhomogeneities during the inflationary era in the early universe}.
 \label{fig2}}}}
\end{figure}

The EBU model of spacetime represents this situation(see Figure 2):
time progresses, events take place, and history is shaped. This is
represented through a growing spacetime diagram, in which the past
is represented as a usual block universe, but now existing only from
the start of space time up to the ever-changing surface representing
the present. Even the nature of future spacetime, along with the
physical events that occur in it, is uncertain; unlike the past, the
future does not yet exist, it is just a potentiality; hence it is
not represented in the diagram as part of the presently existing
spacetime. The passing of time marks the change from indefinite (not
yet existing) to definite (having come into being); the present
marks the instant at which we can act and change reality. Space time
grows as time inexorably evolves: at each new instant every previous
present has become part of the past \cite{Ell06}.

The proposed view is thus that spacetime is continually extending to
the future as events develop along each world line in a way
determined by the complex of causal interactions; these shape the
future, including the very structure of spacetime itself.  The EBU
continues evolving along every world line until it reaches its final
state, resulting in an unchanging Final Block Universe at the end of
time. One might say that then time has changed into eternity. It is
this Final Block Universe that is usually represented in spacetime
diagrams; but it only exists when time has everywhere run its
course.

\subsection{The paradox}\label{sec:paradox}
 This model of spacetime is obviously far more in
accord with our
 daily experience than the standard Block Universe picture; indeed
 everyday data, including the apparent passage of time involved in
 carrying out every single physics experiment, would seem to
 decisively choose the EBU over the Block Universe. The evidence seems abundantly clear.
 Why then do some physicists prefer the latter? If the scientific method
 is to abandon a theory when the evidence is against it, why do
 some hold to it?

This counter viewpoint is put succinctly by Sean Caroll in a
blog:
\footnote{http://blogs.discovermagazine.com/cosmicvariance/2011/09/01/ten-things-everyone-should-know-about-time/}

\begin{quote}
``\emph{The past and future are equally real. This isn't completely
accepted, but it should be. Intuitively we think that the `now' is
real, while the past is fixed and in the books, and the future
hasn't yet occurred. But physics teaches us something remarkable:
every event in the past and future is implicit in the current
moment. This is hard to see in our everyday lives, since we're
nowhere close to knowing everything about the universe at any
moment, nor will we ever be - but the equations don't lie. As
Einstein put it, `It appears therefore more natural to think of
physical reality as a four dimensional existence, instead of, as
hitherto, the evolution of a three dimensional existence.'} ''
\end{quote}
But the question is which equations, and when are they applicable?
As emphasized so well by Eddington (\cite{Edd28}:246-260), our
mathematical equations representing  the behaviour of macro objects
are highly abstracted versions of reality, leaving almost all the
complexities out. The case made in \cite{Ell11a} is that when true
complexity is taken into account, the unitary equations leading to
the view that time is an illusion are generically not applicable
except to isolated micro components of the whole. The viewpoint
expressed by Carroll supposes a determinism of the future that is
not realised in practice: \emph{inter alia}, he is denying the
existence of quantum uncertainty in the universe we experience. But
physics experiments show that uncertainty to be a well-established
aspect of the universe \cite{FeyLeiSan65,GreZaj06},
and it can have macroscopic consequences in the real world, as is
demonstrated by the historic process of structure formation
resulting from quantum fluctuations during the inflationary era
\cite{Dod03}. These inhomogeneities were not determined until the
relevant quantum fluctuations had occurred, and then become
crystalized in classical fluctuations; and they were unpredictable,
even in principle.

Actually, the EBU proposal does not contradict the first part of the
Einstein quote given in Carroll's blog. The core issue not touched
on in that quote is where lies the future boundary of the
4-dimensional spacetime advocated by Einstein. In the usual Block
Universe picture, it is taken to be at the end of time. In the EBU,
it corresponds to the ever-changing present time.

The prime issue arising is that the spacetime view of special
relativity denies the existence of any preferred time slices,
whereas the claimed existence of the present in the EBU is certainly
a preferred time surface (at each instant, it is the future boundary
of the 4-dimensional spacetime). I will deal with this objection in
the following section, after first looking at common physics views
of the passage of time in the rest of this section. An array of
further arguments for the claim ``time is an illusion'' have been
made by philosophers and physicists: they are conveniently
summarized in the Spring, 2012 special issue of \emph{Scientific
American} \cite{SciAm12}\footnote{see also
\cite{Ish92,HugVisWut12,And12} for recent reviews of these issues,
with references.}. I will turn to them in \S \ref{sec:other}. I then
consider the way the Block Universe view relates to theories of the
mind (\S \ref{sec:mind}): a key problem for that view. Next I
consider how the EBU picture may be altered when one takes quantum
issues into account (\S \ref{sec:quant}), and point out how it
relates to the arrow of time issue (\S \ref{sec:arrow}) and solves
the chronology protection question (\S \ref{sec:chron}). Finally I
reflect on the nature of time in relation to the EBU proposal (\S
\ref{sec:close}).

\subsection{The classical physics of the passage of time}
There are no problems with the existence or passage of time in
standard physics textbooks on classical mechanics, see for example
\emph{The Feynman Lectures in Physics} \cite{FeyLeiSan63}.

\subsubsection{Reversible Dynamics}\label{sec:class}
A standard example is a Simple Harmonic Oscillator (`SHO') with
equation of motion
\begin{equation}
    \label{eq:2-1}
F = m \frac{d^2q}{dt^2} = -k q
\end{equation}
and solution
\begin{equation}
    \label{eq:2-2}
q(t) = A \cos (\omega t - \phi), \,\, p(t) = m \frac{dq}{dt} = - m A
\omega \sin (\omega t - \phi)
  \end{equation}
where $\omega := \sqrt{\frac{k}{m}}$. As time evolves, the
oscillator oscillates, with its state at time $t$ given by \{$q(t),
p(t)$\}. Standard texts discussing the SHO do not question the
existence or flow of time.

More generally for dynamical systems with
$N$ variables $x_i$:
\begin{equation}
    \label{eq:2-1a}
dx_i/dt = f(x_j)\,\,\,(i,j = 1-N)
\end{equation}
the solution $ x_i(t)$ represents how the variables change as time
flows steadily on.\footnote{The SHO is of this form if $x_i = x$,
$x_2 = p$.} In these cases knowledge of the state at any time $t_0$
enables deduction of the state at all earlier and later times; the
system is time-reversible and predictable, and evolves with time
according to this equation (indeed the very purpose of these
equations is to predict this time evolution).
A particular case is Hamiltonian dynamics where
\begin{eqnarray}
    \label{eq:Ham1}
dp_i/dt &=& - \frac{\partial}{\partial q_i}{\cal H}(q_j(t),p_j(t)),\\
\label{eq:Ham2}dq_i/dt &=& + \frac{\partial}{\partial p_i}{\cal H}(q_j(t),p_j(t)),
\end{eqnarray}
$(i,j = 1-N)$. There may be constraints, perhaps involving first and second spatial derivatives:
\begin{equation}
    \label{const}
C_m(p_i,q_j,q_{j,k},q_{j,kl}) = 0
\end{equation}
$(m = 1-M)$ where $q_{j,k} := \partial q_j/\partial dx^k$. Then these must be preserved under the time evolution:
\begin{equation}
    \label{const1}
\{(\ref{eq:Ham1}),(\ref{eq:Ham2})\} \,\,\,\Rightarrow\,\,\, d C_m/dt = 0.
\end{equation}
Then provided the constraints are initially satisfied at time $t_0$, the past and the future are uniquely determined for some time interval $[T_-,T_+]$ containing $t_0$:
\begin{equation}
    \label{evolve}
\{(p_i(t_0), q_i(t_0)): C_m(t_0) = 0\} \,\,\,\Rightarrow\,\,\, \{(p_i(t), q_i(t)):\,\, T_-<t<T_+\}.
\end{equation}
The time development of the system is given by these equations. Three comments:\\

\textbf{Explicit time dependence}: The case ${\cal H}(q_j(t),p_j(t),t)$ where $\partial {\cal H}/\partial t \neq 0$ breaks time translation invariance and explicitly invokes preferred times in the dynamics. I exclude this case in what follows.\\

\textbf{Limited prediction times:} generically one or both of $T_-$, $T_+$ will be finite \cite{HawEll73}. Except for comments on the chronology protection question (\S \ref{sec:chron}), I will not consider such global issues here.\\

\textbf{First integrals}: for any function $f(q_i,p_i,t)$ equations (\ref{eq:Ham1}),(\ref{eq:Ham2})  imply the time derivative
\begin{eqnarray}
\label{evolve1}
df(q_i,p_i,t)/dt &=& (\partial f/\partial q_i) (dq_i/dt) + (\partial f/\partial p_i) (dp_i/dt) + \partial f/\partial dt \\
&=& \left(\frac{\partial f}{\partial q_i} \frac{\partial {\cal H}}{\partial p_i} - \frac{\partial {\cal H}}{\partial q_i} \frac{\partial f}{\partial p_i}\right) + \partial f/\partial dt.
\end{eqnarray}
Applying this to the Hamiltonian ${\cal H}$ itself,
\begin{eqnarray}
\label{evolve1}
d{\cal H}(q_i,p_i)/dt =
\left(\frac{\partial \cal H}{\partial q_i} \frac{\partial {\cal H}}{\partial p_i} - \frac{\partial {\cal H}}{\partial q_i} \frac{\partial \cal H}{\partial p_i}\right) = 0,
\end{eqnarray}
so ${\cal H}$ is the conserved energy (related to time translation invariance of the dynamics); in simple cases with kinetic energy $T(p): = \frac{1}{2}p^2$ and potential energy $V(q)$, \begin{eqnarray}
 {\cal H}(p(t),q(t)) = T(p) + V(q) = const =: E . \label{eq:energy} \end{eqnarray}
 This constant relation does not imply there is no time evolution taking place: it means there is a first integral of that evolution.
 If there is no conserved energy, the Hamiltonian description \{(\ref{eq:Ham1}), (\ref{eq:Ham2})\} does not apply. This will be the case whenever  dissipative processes take place and affect the dynamics at the chosen scale of description; this occurs in a great many cases in both macro and micro physics \cite{Ell11}.

\subsubsection{Irreversible Dynamics}\label{sec:irrev}
 In general, friction effects mean we have an inability
to retrodict if we lose information below some level of coarse
graining. The simplest example is a block of mass $m$ sliding on a
plane, slowing down due to constant limiting friction $F = - \mu R$
where $\mu$ is the coefficient of friction and $R = m g$ is the
normal reaction, where $g$ is the acceleration due to gravity
\cite{Spi67}. The motion is a uniform deceleration; if we consider
the block's motion from an initial time $t = 0$, it comes to rest at
some later time $t_* > 0$. For $t < t_*$ the velocity $v(t)$ and
position $x(t)$ of the object are given by
  \begin{equation}
    \label{eq:2-3}
v_1(t) = -  \mu g t  + v_0, \,\,\,   x_1(t) = -  \frac{1}{2} \mu g
t^2 + v_0 t + x_0
  \end{equation}
where $(v_0, x_0)$ are the initial data for $(v,x)$ at the time $t =
0$. This expression shows that it comes to rest at $t^* = \mu
g/v_0$. For $t > t^*$, the quantities $v$ and $x$ are given by
  \begin{equation}
    \label{eq:2-4}
v_2(t) = 0,  x_2(t) = X  \, (constant),
\end{equation}
where $X := -  \frac{1}{2} \mu g t_*^2 + v_0 t_* + x_0$.

The key point now is that from the later data (\ref{eq:2-4}) at any
time $t > t_*$ you cannot determine the initial data $(v_0, x_0)$,
nor even the time $t_*$ when the object came to rest, thus you
cannot reconstruct the trajectory (\ref{eq:2-3}) from that later
data. You cannot even tell if the block came from the left or the
right. The system is no longer time reversible or predictable. The
direction of time is uniquely determined by the way it came to a
halt (it cannot spontaneously start moving; note that the rest frame
implied here is that defined by the table on which the block
slides). That coming to rest was an event that took place in time;
time does not cease after it comes to rest, it is the block's motion
that ceases then. At macro scales, reversibility does not hold, nor
is the motion predictable in both directions of time; the dynamics is not Hamiltonian.

A standard response is, if we knew how the surface was heated, we
could work out where the block cane from and when it arrived. But
the heat dissipates away and vanishes into thermal fluctuations (at
a small enough scale, into quantum fluctuations); that record is
soon irretrievably lost. The claim that ``every event in the past
and future is implicit in the current moment'' soon ceases to be
true.

\subsection{The quantum physics of the passage of
time}\label{sec:quant}

In the last example, our inability to predict is associated with a
lack of detailed information. So if we fine-grained to the smallest
possible scales and collected all the available data, could we then
determine uniquely what is going to happen? No, we can't predict to
the future in this way because of foundational quantum uncertainty
relations(see e.g. \cite{Fey85,Pen89,Ish97}). We cannot predict
precisely when a nucleus will decay or what the velocity of the
resultant particles will be, nor can we predict precisely where a
photon or electron in a double-slit experiment will end up on the
screen.  This unpredictability is not a result of a lack of
information: it is the very nature of the underlying physics.

There are two kinds of quantum evolution \cite{Pen89,Ish97,Pen04}. First there is the unitary
Schrodinger evolution
 \begin{equation}
i\hbar \frac{\partial \ket\psi}{\partial t} = \hat{H} \ket\psi
\label{wave}
 \end{equation}
where $\ket{\psi(t)}$ is the wave function and $\hat{H}$ the
Hamiltonian operator. This equation determines its evolution: time
occurs here in the same way as in classical physics, though the
meaning of the relevant variable is quite different. Time rolls on
and the state vector evolves; therefore probabilities change with
time.

Second there is wave function reduction, associated with both state
vector preparation and measurement events. Consider the wave
function or state vector $\ket{\psi(x)}$. The basic expansion
postulate for quantum mechanics is that before a measurement is made,
$\ket\psi$ can be written as a linear combination of eigenstates
 \begin{equation}
 \ket{\psi_1} = \sum_n c_n\ket{u_n(x)}, \label{wave1}
 \end{equation}
where $u_n$ is an eigenstate of some observable $\hat A$ (see e.g.
\cite{Ish97}: 5-7).  Immediately after a measurement is made at a
time $t=t^*$, however, the wavefunction is found to be in one of the
eigenstates:
 \begin{equation}
 \ket{\psi_2} = c_Nu_N(x) \label{collapse}
 \end{equation}
for some specific index $N$.  The data for $t< t^*$ do not determine
$N$; they merely determine a probability for the outcome $N$ through
the fundamental equation
 \begin{equation}
p_N = c_N^2.  \label{prob}
 \end{equation}
One can think of this as due to the probabilistic time-irreversible
reduction of the wave function\\
\begin{equation}
 \ket{\psi_1} = \sum_n c_n\ket{u_n(x)} \hspace{1 cm}
 \longrightarrow \hspace{1 cm}\ket{\psi_2} = c_Nu_N(x)
\end{equation} \\
(\cite{Pen89}: 260-263). This is the event where the uncertainties
of quantum theory become manifest, as the indeterminate future makes
a transition to the determined past (up to this time the evolution
is determinate and time reversible). Invoking a many-worlds
description (see e.g. \cite{Ish97}) will not help determine the
specific outcome: in the actually experienced universe in which we
make the measurement, $N$ is unpredictable, as confirmed by
experiment. The specific experimental outcome (\ref{collapse}) that
will be measured by an observer to occur at a later time is not determined by the
Everett hypothesis.\footnote{This assumes that in Figure 1, \emph{all} those
possible paths in fact occurred; but we experience only one specific path.}

Thus the initial state (\ref{wave}) does not uniquely determine the
final state (\ref{collapse}); and this is not due to lack of data,
it is due to the foundational nature of quantum interactions. You
can predict the statistics of what is likely to happen but not the
unique actual physical outcome, which unfolds in an unpredictable
way as time progresses; you can only find out what this outcome is
after it has happened. Furthermore, in general the time $t*$  is
also not predictable from the initial data: you don't know when
`collapse of the wave function' (the transition from (\ref{wave}) to
(\ref{collapse})) will happen (you can't predict when a specific
excited atom will emit a photon, or a radioactive particle will
decay).

We also can't retrodict to the past at the quantum level, because
once the wave function has collapsed to an eigenstate we can't tell
from its final state what it was before the measurement. You cannot
retrodict uniquely from the state (\ref{collapse}) immediately after
the measurement takes place, or from any later state that it then
evolves to via the Schrodinger equation at later times $t > t*$,
because knowledge of these later states does not suffice to
determine the initial state (\ref{wave}) at times $t < t*$: the set
of quantities $c_n$ are not determined by the single number $a_N$.

This process takes place all the time as physical events occur and
have classical outcomes (in photosynthesis in plants and in
nucleosynthesis in the early universe, for example); it is not
necessarily associated with a measuring apparatus or the mind of an
experimenter. However it is time-irreversible, causing information
loss, and so is not describable by any unitary evolution.
The classical world would not exist if this did not happen as an
ongoing unfolding process in time.

The fact that such events happen at the quantum level does not
prevent them from having macro-level effects. Many systems can act
to amplify them to macro levels, including photomultipliers  (whose
output can be used in computers or electronic control systems).
This amplification is what occurred when cosmic rays  - whose emission
is subject to quantum uncertainty - caused genetic damage in the
distant past, resulting in new phenotypes occurring \cite{Per91}.
The specific outcome that actually occurred was determined as it
happened, when quantum emission of the relevant photons took place.
Any specific emission event (a photon emission time and trajectory)
was not determined by the priori quantum state, so any consequent
damage to a specific gene in a particular cell at a particular time
and place cannot be predicted even in principle.\footnote{This
damage is not trivial: see \cite{ScaWheWil01}.} Consequently the
specific evolutionary outcomes for life on Earth (the existence of
dinosaurs, giraffes, humans) cannot be uniquely determined by causal
evolution from detailed data at the start of life on Earth.

\section{A problem: Surfaces of change}\label{sec:surf}
The problem however is the claimed unique status of ``the present''
in the EBU - the surface where the indeterminate future is changed
to the definite past at any instant. In this section I propose that
there are indeed such preferred surfaces in all realistic general
relativity models of space-time.\\

It is a fundamental feature of Special Relativity that simultaneity
is not uniquely defined, it depends on the state of motion of the
observer; and one presumes that in some fundamental sense the
present must be regarded as a surface of constant time. What is past
and future elsewhere depends on one's motion, hence the block
universe model is natural: it is the only way a spacetime model can
incorporate this lack of well defined surfaces of instantaneity. For
different observers at a event $P$, different surfaces of
simultaneity will designate different events $Q$ on a distant world
line $\gamma_1$ as simultaneous with $P$ \cite{EllWil00}; the only
resolution is that they are all simultaneous with $P$, hence time is
an illusion.

However there are two fundamental points to be made here that
completely change the picture. First, the physical events that shape
how things evolve are based on particle interactions, and take place
along timelike or null world lines, not on spacelike surfaces, which
are secondary. The concept of simultaneity is only physically
meaningful for neighboring events; it has no physical impact for
distant events, it is merely a theoretical construct we like to make
in our minds. What we think is instantaneous
makes no difference to our interaction with a vehicle on Mars. What
is significant is firstly what happens over there, secondly what
happens here on Earth, and, thirdly the signals between us.
Simultaneity
does not enter into it.

What really matters is proper time $\tau$ measured along
timelines $x^i(v)$, determined from the metric tensor $g_{ij}(x^k)$
by the basic formula \cite{HawEll73,EllWil00}
\begin{equation}
\tau = \int \sqrt{-ds^2} = \int \sqrt{-g_{ij}\,
(dx^i/dv)\,(dx^j/dv)} \,\, dv \label{proper_time}
\end{equation}
Indeed this is the reason why the metric tensor is central to
relativity theory: as well as determining which lines are null lines
($d\tau = 0$ all along the curve), it determines proper time along
timelike world lines. Natural surfaces of constant time are given by
this integral since the start of the universe. Thus we can propose
that
\begin{quote}
\textbf{The present}: \emph{The ever-changing surface $S(\tau)$
separating the future and past - the `present' - at the time $\tau$
is the surface $\{\tau = constant\}$ determined by the integral
(\ref{proper_time}) along a family of fundamental world lines
starting at the beginning of space time}.
\end{quote}
(if the universe existed forever we have to start at some
arbitrarily chosen `present' time $\{\tau_0 = const\}$, which we
assume exists, and integrate from there).

But is this well defined, given that there are no preferred
world-lines in the flat spacetime of special relativity? The second
fundamental feature is that it is general relativity that describes
the structure of space time, not special relativity. Gravity governs
space-time curvature \cite{HawEll73}, and because there is no
perfect vacuum anywhere in the real universe (\emph{inter alia}
because cosmic blackbody background radiation permeates the
Solar System and all of interstellar and intergalactic space
\cite{Dod03}), space time is nowhere flat or even of constant
curvature; therefore there are preferred timelike lines everywhere
in any realistic spacetime model \cite{Ell71}. The special
relativity argument does not apply.

A unique geometrically determined choice for fundamental worldlines
is the set of timelike eigenlines $x^a(v)$ of the Ricci tensor (they
will exist and be unique for all realistic matter, because of the
energy conditions such matter obeys \cite{HawEll73}). Their
4-velocities $u^a(v) = dx^a(v)/dv$ satisfy
\begin{equation}\label{lines}
 T_{ab}u^b = \lambda_1 u_a \,\,\Leftrightarrow \,\, R_{ab}u^b = \lambda_2 u_a
 \end{equation}
 where the equivalence follows from the Einstein field equations.
Thus we can further propose that

\begin{quote}
\textbf{Fundamental world lines}: \emph{the proper time integral
(\ref{proper_time}) used to define the present is taken along the
world lines with 4-velocity $u^a(v)$ satisfying (\ref{lines}).}
\end{quote}
In effect this is the proper time comoving gauge used in
perturbation theories: it will of course give the usual surfaces of
constant time in the standard
Friedmann-Lema\^{i}tre-Robertson-Walker (FLRW) cosmologies.

Two key issues arise  regarding this proposal:
\begin{itemize}
  \item \textbf{What about general covariance and local Lorentz
invariance}? These are symmetries of the general theory, not of its
solutions. Interesting solutions break the symmetries of the theory:
this is not surprising, as we know that broken symmetries are the
key to interesting physics \cite{And72}.
  \item \textbf{What about simultaneity}? In general these surfaces are not
related to simultaneity as determined by radar \cite{EllWil00};
indeed this is even so in the FLRW spacetimes (where the surfaces of
homogeneity are generically not simultaneous, according to the radar
definition \cite{EllMat85}). The flow lines are not necessarily
orthogonal to the surfaces of constant time; indeed they may have
non-zero vorticity and acceleration as well as shear and expansion,
so there may be no surfaces orthogonal to the flow lines
\cite{Ell71}. More than that: the surfaces determined in this way
are not even necessarily spacelike, in an inhomogeneous spacetime.
\end{itemize}
The latter feature means that the there may possibly be a \emph{time horizon}: a null boundary where
these surfaces make a transition from spacelike to timelike. This will of course only happen for very large
gravitational fields such as occur in black holes, indeed these surfaces may well usually coincide with an event horizon.
The initial value problem will then be very different when based in data in these surfaces; however  even if these
surfaces become timelike (necessarily then being null in some places), data on them will still determine the spacetime in their
future and past Cauchy development, up to intersections of this development with surfaces where the outcome is already determined.
The physics of time then will be quite different than usual: this needs investigation. It could relate to the black hole information paradox.\\

\textbf{In summary}: While the general coordinate invariance invoked
in general relativity theory might be thought to proclaim there are
no preferred such surfaces, in any particular solution this is not
the case - there will be preferred timelike lines in any
realistic cosmological solution. The result will be existence of a
family of preferred surfaces representing constant proper time
$\tau$ since the start of the universe along these fundamental world
lines. The proposal is that each represents what was the ``present''
at the corresponding time $\tau$, for all times up up to the present
time $\tau_0$ (they don't exist for $\tau > \tau_0$, for that
spacetime is not yet determined). These surfaces are derivative
rather than primary, as they result from the configuration of
fundamental world lines. They will usually not be instantaneous as
determined by radar soundings.

\section{Other Arguments against an EBU}\label{sec:other}
A series of other arguments, both physical and philosophical, have
been deployed in favour of the standard Block Universe picture
\cite{SciAm12}, and hence deny the EBU proposal. In this section I
review them, and argue that none of them are fatal.

\subsection{Categorization problem} A philosophical argument is that
the past, present, and future are exclusive categories, so a single
event can't have the character of belonging to all three. The
counter is as follows:

Suppose E happens at $t_E$.

At time $t_1<  t_E$, E is in future,

      At time $t_1 = t_E$, E is in present,

    At time $t_1 >t_E$, E is in past.

\noindent Its category changes - that is the essence of the flow of
time - so this is a semantic problem, not a logical one. One needs
adequate semantic usage and philosophical categories to allow
description of this change: language usage can't prevent the flow of
time.

\subsection{Not necessary to describe events}\label{ADM}  Davies \cite{Dav12} and
Rovelli \cite{Rov08} claim time does not flow because it's not
needed to describe the relations between relevant variables, which
are all that matter physically. Thus you can always get correlations
between position $p(t)$ and momentum $q(t)$ for a system by
eliminating the time variable: solve for $t = t(q)$ and then
substitute to get $p(t) = p(t(q)) = p(q)$, and time has vanished!
For example in the case of the simple harmonic oscillator (equation
(\ref{eq:2-2})), this gives the SHO phase plane:
\begin{equation}\label{phase_plane}
q^2 + (p/m)^2 = A^2.
 \end{equation}
Thus one can describe system changes by relating component variables
to one another, rather than to a global idea of time; which suggests
nothing happens or changes, they are just correlated.

Yes indeed, one can find this time-independent representation of
what happens.\footnote{This is just the energy integral (\ref{eq:energy}).} But that does not mean that time does not flow, it
just means that the results of times flow are correlations between
relevant variables. That abstraction represents part of what
happens, namely the relation between $p$ and $q$, and omits other
parts, namely the relation to time. One can put time back to get
\begin{equation}\label{phase_plane1}
q(t)^2 + (p(t)/m)^2 = A^2, \,\,q(t) = A \cos (\omega t - \phi)
 \end{equation}
and the point representing the system moves along the flow lines as
time changes. The first model leaves out part of what is happening:
that does not mean it does not happen, it just means it's  a partial
model of reality, including some aspects and omitting others. It leaves out the way that
the continually changing correlations flow smoothly one after another in a continuous ongoing way.

\subsection{Rates of change}\label{sec:rate}
A key question is, ``What determines the rate of flow of time?'', or
``How fast does time pass?'' Davies and others suggest there is no
sensible answer to this question. In contrast, I claim that the
answer is given by equation (\ref{proper_time}), which determines
proper time $\tau$ along any world line. This is the time that will
be measured along that world line by any perfect clock
\cite{HawEll73,EllWil00}; real world clocks - oscillators that obey
the Simple Harmonic equation - are approximations to such ideal
clocks, and it is the relation between such clocks and other
physical events that measures the passage of time.\\

\textbf{The preferred time parameter} The whole edifice of physics is built on the assumption that we can
build such clocks to a good approximation, giving a time parameter
$\tau$ that appears equally in all dynamical equations of physics:
Newton's equation of motion \cite{FeyLeiSan63}, Maxwell's equations
\cite{FeyLeiSan64}, the Schr\"{o}dinger equation \cite{FeyLeiSan65},
General Relativity expressed in a 1+3 covariant formalism
\cite{Ell71}, and so on; because of this, a lot of money is spent on
building idealized clocks that are understood to be more accurate
than any previous clock\footnote{See e.g.
http://en.wikipedia.org/wiki/Clock; accurate navigation for example
requires accurate timekeeping \cite{Sob95}, which is thus a core
feature of GPS systems.}. Standard physics would not work if you
needed a different time parameter in each of these equations.
Special and general relativity identify that time as proper time
(\ref{proper_time}) along timelike worldlines.

Given such clocks, the rate of change with time of any variable
$f(\tau)$ along a world line is given by
 \begin{equation}
                      f' =  df/d\tau.
 \end{equation}
Then choosing $f(\tau) = \tau$, the answer to the question posed is
that \emph{the rate of change of time is unity}:
\begin{equation}
                      \tau' =  d\tau/d\tau = 1.
 \end{equation}
In other words, through (\ref{proper_time}) the rate of change of
time in any particular coordinate system is determined by the metric
tensor. Using normalised comoving coordinates  with $u^a =
\delta^a_0$ and the time parameter $v$ chosen as $\tau$
\cite{Ell71}, $g_{00} = -1$  and (\ref{proper_time}) becomes
\begin{equation}
\tau = \int d\tau \label{proper_time1}
\end{equation}
The relative flow of time along different world lines may be
different: that is the phenomenon of time dilation, caused by the
varying gravitational potentials represented by the metric tensor
\cite{Tho95}. But this does not mean it is not well defined along
each world line.\\

\textbf{The metric evolution} So if the metric tensor determines proper time, what determines the
metric tensor? The Einstein field equations, of course
\cite{HawEll73}. These can be expressed in many ways, for example a
1+3 covariant formalism \cite{Ell71}, a tetrad formalism
\cite{Ell67}, or the ADM formalism \cite{ADM}. Following the ADM
approach, the first fundamental form (the metric) is represented as
\begin{equation}
ds^2 = (- N^2+ N_i N^i) dt^2 + N_i dx^idt+ g_{ij}dx^i dx^j
\end{equation}
where $i,j = 1,2,3$. The lapse function $N(x^\alpha)$ and shift
vector $N_i(x^\beta)$ represent coordinate choices, and can be
chosen arbitrarily; $g_{ij}(x^\alpha)$ is the metric of the 3-spaces
$\{t = const\}$. The second fundamental form is
\begin{equation}
\pi_{ij} = n_{i;j}\,
\end{equation}
where the normal to the surfaces $\{t = const\}$ is $n_i =
\delta_i^0$; the matter flow lines have tangent vector $u^i =
\delta^i_0$ (which differs from $n^i = g^{ij}n_j$ whenever $N_i \neq
0$, cf \cite{KinEll73}). The field equations for $g_{ij}$ are as
follows (where 3-dimensional quantities have the prefix $(3)$): four
constraint equations
\begin{eqnarray}
{}^{(3)}R &+& \pi^2-\pi_{ij}\pi^{ij} = 16 \pi \rho_H,\label{ADM5}\\
 R^{\mu}&:=&
-2\pi^{\mu j}_{\,\,\, |j} = 16 \pi T^{\mu}_0 \label{ADM3}
\end{eqnarray} where $``|j''$ represents the covariant derivative in the 3-surfaces, and
twelve evolution equations \begin{eqnarray}
\partial _ t g_{ij} &=& 2Ng^{-1/2}(\pi{ij} - \frac{1}{2}g_{ij}\pi) + N_{i|j} +
N_{j|i}, \label{ADM1}\\
\partial_t\pi_{ij} &=& - Ng^{-1/2} ({}^{(3)}R_{ij} - \frac{1}{2}g_{ij} {}^{(3)}R)
+ \frac{1}{2}Ng^{-1/2}g_{ij}(\pi_{mn}\pi^{mn} -\frac{1}{2}\pi^2) -
2Ng^{-1/2}(\pi^{im}\pi_m^{\,\,\,j} \nn  \\  &-&
\frac{1}{2}\pi\pi^{ij}) + \sqrt{g}  (N^{|ij} - g^{ij}
N^{|m}_{\,\,\,|m}) + (\pi^{ij}N^m)_{|m} - N^i_{|m}\pi^{mj} - N^{j}_{
|m}\pi^{mi} + 16 \pi \hat{T}_{ij}. \label{ADM2}
\end{eqnarray}
Equations of state for matter must be added, and the matter
conservation equations $T^{ab}_{\,\,\,\,;b} = 0$ satisfied (as is
required for consistency of the evolution equations). Then
(\ref{ADM1}) determines the rate of change of the metric
$g_{ij}(x^\alpha)$ relative to the ADM time coordinate; (\ref{ADM2})
determines the rate of change of the geometric source terms
$\pi_{ij}$ occurring in (\ref{ADM1}); the matter equations determine
the rate of change of the matter terms. How this works out in
practice is shown in depth in \cite{Ann01}. Overall this determines
the metric tensor as a function of time, and hence evolution of the
surfaces of constant time as defined above (which are determined by
the metric).

This can be worked out using \emph{any} time surfaces (that is the
merit of the ADM formalism); in particular one can specialise the
time surfaces and flow lines to those defined above (\S
\ref{sec:surf}):
\begin{enumerate}
  \item We choose the 4-velocity to be a Ricci Eigenvector:
\begin{equation}
T^{\mu}_0 = 0 \,\,\Rightarrow\,\, R^{\mu}= -2\pi^{\mu j}_{\,\,\, |j}
= 0,\label{fund1}
\end{equation}
which algebraically determines the shift vector $N_i(x^j)$, thereby
solving the constraint equations (\ref{ADM3});
  \item We determine the lapse function $N(x^i)$ by the condition that
the time parameter $t$ measures proper time $\tau$ along the
fundamental flow lines.
\end{enumerate}
These conditions uniquely determine the lapse and shift (see the
Appendix for details). Then, given the equations of state and
dynamical equations for the matter (in the case of a perfect fluid, equations (45)-(47) on page 45 of \cite{Ann01})
with $S^\kappa = S \delta^\kappa_0$ , equations (\ref{ADM1}),
(\ref{ADM2}) determine the time evolution of the metric in terms of
proper time $\tau$ along the fundamental flow lines; the constraints are conserved
because of energy-momentum conservation. The development
of spacetime with time takes place just as is the case for other
physical fields, with the relevant time parameter being proper time
along the fundamental flow lines. There is no problem
with either the existence or the rate of flow of time.\\

\textbf{Predictability}: Do these equations mean the spacetime development is uniquely determined to the future and the past from initial data? That all depends on the equations of state of the matter content: the relations between the density $\rho_H$ in (\ref{ADM5}), pressure tensor $\hat{T}_{ij}$ in (\ref{ADM2}), and momentum density $T^{\mu}_0$  in (\ref{ADM3}). These quantities depend on the frame chosen, and $T^{\mu}_0$  is zero when we make the choice (\ref{fund1}).

Assuming this choice, define the pressure $p$ and anisotropic stress $\Pi_{ij}$ by
\begin{equation}
p = \frac{1}{3}\, g^{ij}\hat{T}_{ij}, \,\,\Pi_{ij} = \hat{T}_{ij} -  p g_{ij}.
\end{equation}
The future and past will be uniquely determined for simple equations of state such as non-interacting baryons plus radiation, as in standard cosmological models:
\begin{equation}
\rho_H =  \rho_{b} + \rho_{r},\,\, p = p_{b} + p_{r} = \frac{1}{3}\rho_{r},\,\,\Pi_{ij} = 0
\end{equation}
where the energy density conservation equations will determine the time evolution of $\rho_{b}$, $\rho_{r}$. However 
\begin{itemize}
  \item one can have dissipative processes (shear viscosity or bulk viscosity, for example \cite{Ell71}), so the evolution is not time reversible  - a Hamiltonian description does not apply to the matter, and hence also does not apply to the combined (matter, gravity) system, where matter determines space-time curvature;
  \item  one can have an explicitly time dependent equation of state:
\begin{equation}
p = p(\rho_H,\tau), \,\,\Pi_{ij} = \Pi_{ij}(\rho_H,\tau)
\end{equation}
and predictability is no longer the case if the time dependence is not predictable from the available initial data at the relevant scales; again a time-reversible and predictive Hamiltionian description can't be used for the system as a whole.
\end{itemize}
For example one can have a massive body where effectively one has
\begin{equation}
\Pi_{ij}(\tau) = F(\tau)\, \Pi_{ij}(0)
\end{equation}
where $F(\tau)$ represents internal dynamics not visible to the external world\footnote{Bondi's massive objects Tweedledum and Tweedledee incorporated this idea: see Narlikar \cite{Nar} for a description.}; there might be a mechanism
here that is computer controlled via an algorithm embodying a random number generator\footnote{see http://www.random.org/randomness/ for a discussion.}
or based in random signals generated via radioactive decay \cite{rand}. Then as explained above (\S \ref{sec:EBU} and Fig.\ref{fig1}), the initial data do not determine the outcome ($F(\tau$) indicates a causal influence but not a predictable functional relation). One can only determine what will happen as it happens.

So equations (\ref{ADM1}), (\ref{ADM2}) determine the time evolution of the spacetime, but do not guarantee predictability either to the future or the past. That depends on the physics of the matter; if quantum unpredictability gets amplifed to macro scales, the spacetime evolution is intrinsically undetermined till it happens (as mentioned above, this was essentially what happened during the generation of seed inhomogeneities in the inflationary era in the very early universe).\\

\noindent \textbf{Conclusion}: time flows at the rate of one second
per second, with the metric tensor determining what this rate is for
clocks and every other physical system (the choice of units is of
course arbitrary, but can be done in a way that makes sense). The
result is that clock readings and particle motions are correlated in
a way that enable us to reliably predict motions. So yes, such
correlations are fundamental to our experience of the flow of time
(as emphasized by Davies); they are a result of its inexorable
omnipresent continuing flow equally in all physical systems. Can you
change the rate of time? No! Can you stop time? No! Can you reverse
time? No! Like Old Man River, it just keeps flowing on; that is the
primitive expressed in all the time evolution equations of physics
\cite{FeyLeiSan63,FeyLeiSan64,FeyLeiSan65,Ell71} and the related
existence and uniqueness theorems \cite{HawEll73}.

\subsection{Time parameter invariance of  General Relativity}\label{sec:param}
What about the time parameter invariance of General Relativity, as
made manifest in the ADM formalism \cite{ADM,GamPul12}? This has basically
already been dealt with in the discussion above:
\begin{itemize}
  \item The gravitational side of the ADM equations may be
  time-parameter invariant, but the matter side is not, in particular because
  rescaling time changes the value of the kinetic energy $T(p)$ while leaving
  the potential energy $U(x^i)$ unchanged. Hence any
  solutions with matter present (i.e. all realistic solutions) will
  not be time parameter invariant; this is part of the ongoing
  tension between the geometric and matter sides of the Einstein
  Field Equations;
  \item This is part of the broader theme mentioned above: specific
  solutions of the theory have less symmetry than the theory itself;
  this symmetry breaking is a key feature of all realistic solutions
  of the equations of physics \cite{And72}, and in particular cosmological solutions
  (\S\ref{sec:surf});
  \item Proper time (\ref{proper_time}) along fundamental world lines
  (\ref{lines}) provides a preferred
  time parameter in realistic solutions of general relativity theory.
\end{itemize}

\noindent Local physics does indeed have a preferred time parameter:
for example in a Simple Harmonic Oscillator using standard time $t$,
$q(t) = A \cos (\omega t - \phi)$ (see (\ref{eq:2-2})); these cycles
measure time $t$ like a metronome (which is why SHO's are used as
clocks). One can in principle change to an arbitrary time $t'$,
\begin{equation}
t'= t'(t) \Rightarrow t = t (t') \Rightarrow q = A \,\cos\, (\omega
t(t') - \phi),
\end{equation}
for example
\begin{equation} t'= \exp\, H(t) \Rightarrow t =
\exp\,(-Ht') \Rightarrow q =  A\, \cos (\omega\, \exp(-Ht')- \phi)
\end{equation}
hence the regular motion no longer is represented as regular. This
is because one has chosen a peculiar time, which will not correlate
simply with any other physical behaviour either (\cite{GamPul12}:57-60). The sensible choice
of time is that which makes sense of patterns of physics behaviour;
so the maximum sensible variation is $t' = \alpha t  + \beta$ (just
as in the case of the allowed change in affine parameter along
geodesics). One can choose any other reparametrisation of time; but
any except affine transformations of proper time confuse and hide
what is actually happening. One is able to choose proper time, and
does so if one wants to illuminate the physics.

It is non-optimal to examine the dynamics of general relativity
without acknowledging the central role of the metric tensor $g_{ij}$
and resultant proper time (\ref{proper_time}) along world lines -
which are at the core of the physical interpretation of General
Relativity \cite{HawEll73,EllWil00}. You can use a proper time
coordinate $\tau$ in the ADM formalism (as shown above in \S
\ref{ADM}; this is a general way one can solve the problem of time
in ADM dynamics, see \cite{Per97}). That choice ties this time
parameter in to the rest of physics, and in particular to time as
measured by local ideal clocks (such as a cesium atom). Consequently
the flow of time is then characterised by the relation between such
clocks and other physical events, including the gravitational
dynamics represented by the ADM evolution equations (\ref{ADM1}),
(\ref{ADM2}). In this sense time is relational (cf: \cite{GamPul12}:163-166).

In the case of the standard FLRW models of cosmology, the usual
metric is
\begin{equation}\label{FLRW}
ds^2 = - d\tau^2 + a^2(\tau) d\sigma^2
\end{equation}
where $d\sigma^2(x^i)$ is a time-independent 3-space of constant
curvature \cite{Ell71,Dod03}, and $\tau$ is precisely the preferred
time coordinate defined above (\S \ref{sec:surf}), as the flow lines
$u^a = \delta^a_0$ are Ricci flow lines and $\tau$ is proper time
along them (these are normalized comoving coordinates \cite{Ell71}).
Just as in the case of the simple harmonic oscillator, one can
choose arbitrary other time parameters - but it is very perverse to
do so, and this is not done in practice, except for one other common
choice: use of a conformal time parameter $\eta(\tau)$:
\begin{equation}\label{FLRW1}
 \eta = \int d\tau/a(\tau) \,\, \Rightarrow\,\, ds^2 = a^2(\eta)(- d\eta^2 +  d\sigma^2)
\end{equation}
which is a very poor representation of proper time but represents
causal structure very well \cite{HawEll73,EllWil00}. However this
still has the same surfaces of constant time as in (\ref{FLRW}), and
is always acknowledged to be different from the proper time $\tau$
that is fundamental in local physics. One never finds proposals for
a time $t = t(\tau,x^i)$ with $\partial t/\partial xi \neq 0$, which
would have different surfaces of constant time. The usual choice as
in (\ref{FLRW}), agreeing with the proposal made here in \S
(\ref{sec:surf}), ties time $\tau$ in to all the rest of physics,
astronomy, geology, technology, and biology.

\textbf{In conclusion}: Standard physics is based in choice of a
preferred time parameter $\tau$ along matter world lines. General
Relativity both allows such a choice, and can itself be written in
terms of that choice.

\section{Time with an underlying timeless substratum} \label{sec:emerge}
There are a number of proposals for an effective time to emerge
somehow in the context of a timeless substratum. These include the
Mott proposal (\S \ref{sec:Mott}), the Rovelli proposal (\S
\ref{sec:Rovelli}), and proposals based in the the Wheeler-de Witt
equation (\S \ref{sec:WdeW}). In this section I will comment on them
in turn.\\

As a preliminary, I first remark that there are two common themes
cutting across them all:
\begin{itemize}
  \item If an effective time emerges at the macro scale, then however that happens,
  it emerges: and the EBU proposal is then good at macro scales, no matter how
  it relates to a timeless substrate;
  \item All of these approaches are based on unitary Schr\"{o}dinger evolution,
  so none of them effectively tackles the non-unitary evolution associated with both
  state vector preparation and quantum measurements \cite{Ish97}. Hence
  they are omitting a key way that a flow of time takes place at the micro level (\S
  \ref{sec:quant}).\footnote{The many worlds view often associated with
  the Wheeler-de Witt equation proposes to deal
  with measurements, but not with state vector preparation, which is also non-unitary
  \cite{Ell11a}.}
 \end{itemize}
 I regard the latter as a particularly significant problem for all
 such proposals (cf. \cite{Ell11a}).

\subsection{Interaction with the environment}\label{sec:Mott} Mott
\cite{Mot31} and Briggs and Rost \cite{BriRos00,BriRos01} suggest
that the Time Independent Schrodinger Equation (`TISE') is more
fundamental than the Time Dependent Schrodinger Equation
(\ref{wave}) (`TDSE'): indeed that the latter emerges from the
former, the interaction between parts in a timeless whole generates
an effective time characterizing interaction between the parts. This
is summarised in \cite{BriRos01} as follows:
\begin{quote}
\emph{``Following work of Born, in 1931 Mott(3) described the impact
of $\alpha$-particles on atoms by treating both atom and beam
quantum-mechanically with the TISE. Then he showed that for a high
energy beam he could describe its motion classically resulting in a
time-dependent Hamiltonian and TDSE for the atom alone ... time is
entering only from a classical interacting environment .. time
enters the quantum Hamiltonian only when some external system is
approximated by classical behaviour''}.
\end{quote}
Thus this is an emergence of time by a top-down interaction from the
environment.

According to Briggs and Rost \cite{BriRos01}, one can use the TIDE
in the form
\begin{equation}
H\Psi=E\Psi \Leftrightarrow (H_{\cal E}+ H_{\cal S}+ H_{\cal I})\Psi
= E\Psi
\end{equation}
where ${\cal E}$ represents the environment, ${\cal S}$ the system
and ${\cal I}$ their interaction. Through a kinetic energy term, the
interaction Hamiltonian somewhat mysteriously introduces an
effective time into the wave function for the system. If one accepts
this proposal, it is a way that an effective time variation is
induced at the quantum level due to top-down effects from the
environment - a proposal that is in consonance with the broad
suggestions of the effectiveness of top-down effects in quantum
physics presented in \cite{Ell11a}. Once this has occurred, one has
an effective EBU situation at the micro as well as the macro level.

\subsection{Get it by coarse graining?} \label{sec:Rovelli} By
contrast Rovelli \cite{Rov08} suggests time can emerge in a
bottom-up way from a timeless substrate as a thermodynamic variable.
I find it difficult to see how any process of coarse graining a
static state can introduce time, but in any case this proposal faces
two other problems.

First, it is based in the idea of equilibrium distribution:
``\emph{Whatever the statistical state $\rho$ is, there exists
always a variable $t_\rho$, measured by the thermal clock, with
respect to which the system is in equilibrium and physics is the
same as in the conventional nonrelativistic statistical case}!''
However equilibrium is a state that emerges through molecular
collisions: if there is no time there will be no such collisions,
and no reason whatever to assume that equilibrium is the most
probable state of the system. This proposal embodies a hidden
assumption that time already exists.

Second, it is based on the underlying symmetries of Hamiltonian
dynamics: ``\emph{mechanics does not single out a preferred
variable, because all mechanical predictions can be obtained using
the relativistic hamiltonian H, which treats all variables on equal
footing}''. But this symmetry applies particularly to the direction
of time: Hamiltonian dynamics (\ref{eq:Ham1}), (\ref{eq:Ham2}) has a time-reversal symmetry
\begin{equation}
x \rightarrow x, \,\,t \rightarrow -t, \,\,q \rightarrow q, \,\, p
\rightarrow -p,
\end{equation}
hence like every other proposal for the purely bottom up emergence
of time, it has problems determining the arrow of time. An initial reaction
to any such proposal is that coarse graining from micro to macro
scales convincingly results in an arrow of time, as shown
beautifully by Boltzmann's H-Theorem (\cite{Zeh07}:43-48), resulting
from the fact that random motions in phase space takes one from less
probable to more probable regions of phase space
(\cite{Pen04}:686-696; \cite{GemMicMah04}:43-47; \cite{Pen11}:9-56).
Hence one can show that entropy increases to the future; the second
law of thermodynamics at the macro level emerges from the coarse
grained underlying micro theory. The quantum theory version of this
result is the statement that the density matrix of open system
evolves in a time asymmetric manner, leading to an increase in
entropy (\cite{BrePet06}:123-125).

But this apparent appearance of an arrow of time from the underlying
theory is an illusion, as the underlying theory is time symmetric,
so there is no way an arrow of time can emerge by any local coarse
graining procedure. Indeed the derivation of the increase of entropy
in Boltzmann's H-Theorem applies equally to both directions of time
(swap $t \rightarrow -t$, the same derivation still holds). This is
\emph{Loschmidt's paradox} (\cite{Pen89}: Fig 7.6;
\cite{Pen04}:696-699; \cite{Pen11}):
\begin{quote}
\textbf{Time symmetry of the H-Theorem}: \emph{Boltzmann's H-theorem
predicts entropy will increase to both the future and the past}.
\end{quote}
The same will apply to the quantum theory derivation of an increase
of entropy through evolution of the density matrix
(\cite{BrePet06}:123-125, \cite{GemMicMah04}:38-42, 53-58): it
cannot resolve where the arrow of time comes from, or indeed why it
is the same everywhere. The latter is a key question for any local
proposal for determining the arrow of time:
\begin{quote}
\textbf{The arrow of time locality issue}:\emph{ If there is a
purely local process for determining the arrow
  of time, why does it give the same result everywhere}?
\end{quote}
We are unaware of any contradictions as regards the direction of the
arrow of time in the universe around us, either locally (time does not run backwards anywhere
in Earth) or astronomically (irreversible process in distant
galaxies seem to run in the same direction of time as here
\cite{Ree95}). Some top-down coordinating mechanism is called for to
guarantee the future direction of time will be the same everywhere;
that is lacking in any purely bottom-up proposal, which is by its nature based in local interactions only.

Kupervasser \emph{et al} \cite{KupNikZla12} suggest that interaction
between two subsystems with a different arrow of time will cause a
decay towards a universal direction for the arrow of time. This is a
very interesting claim, but has two problems: first, how can a
coherent interaction take place at an interface where the direction
of time is different on the two sides? It seems \emph{a priori} that
paradoxical behaviour will abound, as closed causal loops will necessarily occur
there. And if this does work despite these problems, then one has to
show that there is sufficient time available since the start of the
universe that all domains opposite to the dominant one get coerced
to join the majority; effective causal horizons \cite{HawEll73}
might even prevent this occurring. Kupervasser \emph{et al} do
not put their proposal in the cosmic context, so this too is
unresolved.

\subsection{The Wheeler de Witt equation}\label{sec:WdeW}
A whole literature on the problem of time in quantum cosmology
\cite{Hal03,Bar99,Car10} suggests that an effective time emerges
from a time-independent wave function of the universe determined by
the Wheeler-de Witt equation \cite{Har03,Haw84,IshBut99}. As stated
by Hartle \cite{Har03},
\begin{quote}
``In quantum mechanics, any system - the universe included -- is
described by a wave function $\Psi$. There is a local dynamical law
called the Schr\"{o}dinger equation that governs how the wave
function changes in time:
\begin{equation}\label{Sch_H}
i \hbar \frac{d\ket{\Psi(t)}}{dt} = H \ket{\Psi(t)} \,\,\,\,\,{\rm
(dynamical\, law)}
\end{equation}
Here the operator H, the Hamiltonian, summarizes the dynamical
theory.... the Schr\"{o}dinger equation doesn't make any predictions
itself, it requires an initial condition. This is
\begin{equation}\label{Sch_H0}
\ket{\Psi(0)}\,\,\,\,\,{\rm (initial\, condition)}
\end{equation}
When we consider the universe as a quantum mechanical system, this
initial condition is Hawking's wave function of the universe
\cite{Haw84}.''
\end{quote}
This is the basis of quantum cosmology \cite{GibSheRan03}.

 The problem of time then arises because in the case of General Relativity, where $\ket{\Psi}$ is
 a function of 3-geometries: $\ket{\Psi} = \ket{\Psi(h_{ij})}$, the Hamiltonian is
 such as to lead to the Wheeler-de Witt equation for the wave function of the universe:
\begin{equation}
H  \ket{\Psi} = 0. 
\label{WdeW}
\end{equation}
So by (\ref{Sch_H}), $\ket{\Psi}$ is time independent and the
probabilities for quantum outcomes in the universe, which are
expressed by the wave function $\ket{\Psi}$, are unchanging in time
(this is a simplified sketch; for details, see
\cite{Hal03,And12,HugVisWut12}). Hence time does not pass, the
universe just is. Time is an illusion \cite{Bar99}. A large
literature then tries to show how an effective time can emerge from
this timeless context \cite{Hal03,Bar99,Car10}.
I will make four points.\\

\noindent \textbf{First}, Arnowitt, Deser and Misner write of the
Hamiltonian formalism as follows \cite{ADM}:
\begin{quote}
\emph{Since the relation between $q_{M+1}$ and $\tau$ is
undetermined, we are free to specify it explicitly, i.e., impose a
``coordinate condition''. If, in particular, this relation is chosen
to be $q_{M+1} = \tau$ (a condition which also determines $N$), the
action (2.4) then reduces [to] (2.5) with the notational change
$q_{M+1} \rightarrow \tau$; the non-vanishing Hamiltonian only
arises as a result of this process.}
\end{quote}
This is the choice made above (\S \ref{sec:rate}); the corresponding
Hamiltonian will be non-zero as indicated in this
quote, so (\ref{WdeW}) will not hold (cf. \cite{Per97}).\\

\noindent \textbf{Second}, is there an alternative proposal that
does not lead to (\ref{WdeW})? Yes indeed: unimodular gravity (which
produces a trace free version of the Einstein Field Equations
\cite{Elletal11}) has the same effective gravitational equations as
General Relativity Theory \cite{Wei89} but makes $H\neq 0$ and so
solves the time problem of quantum cosmology. Smolin \cite{Smo09}
states this as follows: ``\emph{Sorkin \cite{Sor94} and Unruh
\cite{Unr89,UnrWal89} have pointed out that unimodular gravity has a
nonvanishing Hamiltonian and hence evolves quantum states in terms
of a global time given by an analogue of the Schrodinger
equation}''. This removes the basis of the problem.

It has other major benefits: as emphasized by Weinberg \cite{Wei89}
and Smolin \cite{Smo09}, it also solves the strong cosmological
constant problem: the  discrepancy factor of at least $10^{70}$
between estimates of the vacuum energy density and the
cosmologically determined value of the cosmological constant
\cite{Dod03}. This resolution  is a crucial need in relating quantum
field theory to general relativity: it is a \emph{sine qua non} for
consistent physics, hence
\begin{quote}
\textbf{Reconciling General Relativity and Quantum Field Theory:
}\emph{evidence from cosmology \cite{Dod03} of the small size of the
cosmological constant strongly favours the trace-free version of the
Einstein Equations over the usual version
\cite{Elletal11}}.\footnote{If one applies Ockham's razor
(``entities must not be multiplied beyond necessity)'' the proposed
multiverse solution is severely disfavored in comparison with this
resolution of the issue, which does not involve an infinitude of
unobservable entities.}\end{quote} Then we can have (\ref{Sch_H})
without (\ref{WdeW}).\\

 \noindent \textbf{Third}, this analysis assumes a
Hamiltonian evolution (\ref{Sch_H}) holds for the wave function of
the universe as a whole at all times. This is sometimes justified by
saying that as there can be no external measuring apparatus for the
universe as a whole to interact with, no measurement or wave
function collapse can take place: only unitary evolution will occur.
However one can advocate an emergent view of higher level laws of
causation from lower level physical interactions \cite{Ell11a}. On
this view, the wave function of the universe $\ket{\psi_U}$ is the
wave function obtained by composition of all its components: it is a
sum of terms of the form
\begin{equation}
 \ket{\psi_{U}} =
 \ket{\psi_{1}} \otimes \ket{\psi_{2}} \otimes ....  \otimes \ket{\psi_{N}}
 \label{productN} \end{equation}
 where $N$ is the number of constituents making up the universe and $\ket{\psi_{i}} $
  the wave function for degrees of freedom of the $ith$ component.
  This will not evolve unitarily if any wave function
collapse takes place for any component anywhere in the universe.
But measurements do
take place and classical physics does emerge. Thus on this view, we
are not entitled to assume existence of a wave function for the
entire universe that always obeys (\ref{WdeW}): this
starting assumption is unjustified, at least at recent times.\\

\noindent \textbf{Finally}, the argument above is centrally based on
the Wheeler de Witt equation, but it is not a tested and proven part
of physics:
\begin{quote}
\textbf{Missing Confirmation}: \emph{we have no observational or
experimental evidence that the Wheeler de Witt equation in fact
describes the evolution of the real universe at any time}.
\end{quote}
 It is an untested extrapolation of known physics, which extrapolation may or may not
represent reality adequately \cite{But10}. Actually one can suggest
that everyday experience strongly suggests it is not true (this is
hinted at in \cite{HugVisWut12}). That is the topic of the next
section.

\section{It's all in the mind}\label{sec:mind}
Barbour \cite{Bar99} tackles the key issue of how the mind can
experience the passing of time in the timeless context of the
Wheeler de Wit equation: how do we reconcile the conclusion that
time is an illusion with the fact we do indeed experience the
passage of time? This section reviews that suggestion, and argues
that it is fatally flawed; and that this issue is a major problem
for all proposals that time is an illusion.\\

Barbour claims \cite{Bar99,But10} there exist records of events that
our brains read sequentially, and so create a false illusion of the
passage of time. Thus brain processes are responsible for illusion
of change. But ``processes'' are things that unfold in time! - there
are no processes unless time flows. You can't perceive a flow of
time unless time flows, because perception is a process that takes
place in time.

The prevalent view of present day neuroscience \cite{KanSchJes00} is
that mental states $\Phi$ are functions of brain states $B$ which
are based in the underlying neuronal states $b_i$, determined by
genetics, chemistry, and physics interactions in the brain, taking
place in the overall physical, social and psychological environment
${\cal E}$. Thus\
\begin{equation}
\Phi = \Phi(B)= \Phi(b_i,{\cal E}).
\end{equation}
If time does not flow in microphysics, in a given unchanging
environment
\begin{equation}\label{eq:nochange}
\{db_i/dt = 0,\, d{\cal E}/dt=0 \}\Rightarrow d\Phi/dt =0:
\end{equation}
mental states cannot evolve, unless they are driven in some
mysterious unspecified way by changes in the the environment: but on
Barbour's argument, that too cannot evolve.

But one thing we do know is that time does flow in our experience
(indeed `knowing' is a key part of that experience!).
 Hence the assumption that time does not flow in the underlying
 microphysics cannot be true: the data proves it to be wrong. If Barbour's view is
correct and no physical events take place, then -- as the brain is
based in physics -- no such record-reading processes can take place.
Rather than showing time is an illusion, I suggest the implication
runs the other way:
\begin{quote}
\textbf{Taking everyday life seriously}: \emph{comparing the
conclusion (`time is an illusion') with the evidence from mental
life, by (\ref{eq:nochange}) the contradiction between them is proof the WdeW
equation (\ref{WdeW}) does not apply to the universe as a whole at
the present time, as proposed by Barbour. }
\end{quote}
If there is a meaningful wave function of the universe
 (perhaps defined by (\ref{productN})), it
does not evolve in a unitary way. The great merit of Barbour's book
is that it takes the Wheeler de Witt equation seriously, and pursues
the implications to their logical conclusion; the evidence from
daily life then shows it to be wrong.

A further set of issues arise as regards the perception of time.
Experienced time $\tau_{exp}$ is a function of proper time $\tau$
but also on the emotional and psychological context. It also has a
minimal resolution $\tau_{min}$ because of the interaction times in
the brain: we cannot distinguish events at smaller times scales
because our senses necessarily average over micro times scales with
a window function $W(\tau_{min})$. But all this is unrelated to the
fundamental issues of the nature of time we are dealing with here:
it is to do with brain functioning. The brain is not necessarily a
good clock, and it will not click over instantaneously: there will
be a finite width to its time resolution. But the fact it works as
it does is evidence of the flow of time in physics, because the
brain is based in physics. It's not just a series of correlations:
it's an ordered sequence of causally related correlations that flow
from each other in an ongoing process enabling mental life.

\section{Taking delayed choice quantum effects into account}\label{sec:quantum}
This paper so far is based on a classical view of physics. The EBU
proposal made so far does not take into account the delayed choice
experiments of quantum theory, which suggest one can in some
circumstances ``reach back into the past'' to affect things there.
This section briefly comments on how one can extend the EBU model to
take this feature into account.\\

One can extend the EBU view to one that takes account of this aspect
of quantum physics through proposing a ``Crystallizing Block
Universe" (CBU), where ``the present" is effectively the transition
region in which quantum uncertainty changes to classical
definiteness \cite{EllRot10}. Such a crystallization, however, does
not take place simultaneously, as it does in the simple classical
picture. Quantum physics appears to allow some degree of influence
of the present on the past, as indicated by such effects as
Wheeler's delayed choice experiments \cite{Whe78,Jacetal07} and
Scully's quantum eraser \cite{Kimetal00} (see the summaries of these
effects in \cite{AhaRoh05,GreZaj06}).

The CBU picture is an extension of the EBU where local events may
lead or lag the overall flow of time, thus allowing some apparent
influences from the future to the past as evidenced in those
experiments. It adequately reflects such effects by distinguishing
the transitional events where uncertainty changes to certainty,
which may in some cases be delayed till after the apparent ``present
time."\\

\noindent I have not here tried to relate the EBU picture to the
issue of entanglement and EPR type of experiments
\cite{AhaRoh05,GreZaj06}. The way those experiments relate to
simultaneity and the flow of time as proposed here is unclear; this
is a topic for future research.

\section{The arrow of time and closed time like lines}
Two closely related problems are the arrow of time problem, and the
issue of closed timelike lines. This section discusses how the EBU
proposal reformulates the first issue, and solves the second.

\subsection{The Arrow of time}\label{sec:arrow}
 As regards the arrow of time problem \cite{Dav74,Zeh07},
if the EBU view is correct, the Wheeler-Feynman prescription for
introducing the arrow of time by integration over the far future
\cite{WheFey45}, and associated views comparing the far future with
the distant past \cite{Pen89,EllSci72,Ell02}, are  invalid
approaches to solving the arrow of time problem, for it is not
possible to do integrations over future time domains if they do not
yet exist. Indeed the use of half-advanced and half-retarded Feynman
propagators in quantum field theory then becomes a calculational
tool representing a local symmetry of the underlying physics that
does not reflect the nature of emergent physical reality, in which
that symmetry is broken.

The arrow of time problem in this EBU context is revisited in a
companion paper \cite{Ell11}. The key point is that
\begin{quote}
\textbf{The direction of time}: \emph{The arrow of time arises
fundamentally because the future does not yet exist: a global
asymmetry in the physics context. The Feynman propagator can only be
integrated over the past, as the future spacetime domain is yet to
be determined.}
\end{quote}
One can be influenced at the present time from many causes lying in
our past, as they have already taken place and their influence can
thereafter be felt. One cannot be influenced by causes coming from
the future, for they have not yet come into being.  The history of
the universe has brought the past into being, which is steadily
extending to the future, and the future is just a set of unresolved
potentialities at present. One cannot integrate over future events
to determine their influence on the present not only because they do
not yet exist, but because they are not even determined at present
(\S \ref{sec:EBU}).

The direction of the arrow of time is thus determined in a
contingent way in the EBU context \cite{Ell11}: it is the direction
of time leading from what already has come into existence (the past)
to the present. Collapse of the quantum wave function is a prime
candidate for a location of a physical solution to the
coming-into-being problem, and manifests itself as a form of
time-asymmetric top-down action \cite{Ell08} from the universe as a
whole to local systems (cf. \cite{Pen89}). A key further feature is
that the initial state of the universe was very special with very
low entropy \cite{Pen11,Car10,Ell11}, allowing complex higher
entropy structures to form later on. But that effects how the arrow
of time works out, leading to the Second Law of Thermodynamics,
rather than its very existence; that is provided by the EBU context.

\subsection{Closed timelike lines: Chronology protection}\label{sec:chron}

A longstanding problem for general relativity theory is that closed
timelike lines can occur in exact solutions of the Einstein Field
Equations with reasonable matter content, as shown famously in the
static rotating G\"{o}del solution \cite{HawEll73}. This opens up
the possibility of many paradoxes, such as  killing your own
grandparents before you
were born and so creating causally untenable situations. \\

It has been hypothesized that a \emph{Chronology Protection
Conjecture} \cite{Haw92} would prevent this happening. Various
arguments have been given in its support \cite{Vis02}, but this
remains an \emph{ad hoc} condition added on as an extra requirement
on solutions of the field equations, which do not by
themselves give the needed protection.\\

The EBU automatically provides such protection, because creating
closed timelike lines in this context requires the undetermined part
of spacetime intruding on regions that have already been fixed. But
the evolving spacetime regions can never intrude into the completed
past domains and so create closed timelike lines through some
spacetime event $P$, because to do so would require the fundamental
world lines to intersect each other either before reaching $P$, or
at $P$. Assuming plausible energy conditions, that would create a
space-time singularity \cite{HawEll73}, because (being timelike
eigenvectors of the Ricci tensor) they are the average flow lines of
matter, and in the real universe, there is always matter or
radiation present: $R_{ab} \neq 0$. The extension of time cannot be
continued beyond such singularities, because they are the boundary
of spacetime.

\begin{quote}
 \textbf{Causality}:\emph{ The existence of closed timelike lines
(\cite{Car10}:93-116) is prevented in an EBU, because if the
fundamental world lines intersect, a spacetime singularity occurs
\cite{HawEll73}: the worldlines are incomplete in the future, time
comes to an end there, and no ``Grandfather Paradox'' can occur.}
\end{quote}
Hence the EBU as outlined above automatically provides chronology protection.

\section{Overall: A more realistic view} \label{sec:close}
\label{sec:time}

This paper has proposed an Evolving Block Universe (EBU)
representation of spacetime which grows with time as events happen.
This final section reviews how it relates to the basic features
usually expected of time, and to some of the surprising features of time
revealed to us by relativity theory.\\

The EBU model recognizes that the nature of the future is completely
different from the nature of the past. The past has taken place and
is fixed, and so the nature of its existence is quite different than
that of the indeterminate future. Uncertainty exists as regards both
the future and the past, but its nature is different in these two
cases. The future is uncertain because it is not yet determined: it
does not yet exist in a physical sense (although it is constrained
in key ways by the current state of things). Thus this uncertainty
has an ontological character. The past however is fixed and
unchanging, because it has already happened, and the times when it
happened cannot be revisited; but our knowledge about it is
incomplete, and can change with time. Thus this uncertainty is
epistemological in nature.\\

\noindent In Newtonian theory, and in ordinary quantum theory, time
is the source of

1.  ordering of events,

2.    duration measured between events,

3.   simultaneity: synchronisation of distant events,

4.    direction of the flow of time,

5.    transition: the fact that time flows,

6.    continuity of the flow of time,

7.    monotonic nature of that flow [it can't reverse or close
up].\\

\noindent But Special Relativity and General Relativity changed
that, with a surprising find \cite{HawEll73}:
\begin{quote}
\textbf{Key discovery 1}: \emph{simultaneity 3. is not fundamental
to time:-   time flows along timelike world lines, proper time along
world lines is the fundamentally preferred time parameter. It is
measured by the spacetime metric, which determines duration 2.}
\end{quote}
Thus simultaneity 3.  is secondary, with no direct physical
consequences. What matters are interactions between distinct
entities; these take place via timelike curves and null geodesics,
not on spacelike surfaces. This potentially puts a major barrier in
the way of the EBU proposal where the flow of time is taken
seriously, but this paper has suggested those barriers  are resolved
by identifying preferred timelike curves and associated spacelike
surfaces in realistic models  of the real universe (\S
\ref{sec:rate}).

Additionally, general relativity made a crucial difference to 7.:
\begin{quote}
\textbf{Key discovery 2}: - \emph{monotonicity 7. is not necessarily
true in a curved spacetime, unless something prevents it (as shown
by G\"{o}del, closed timelike lines are potentially possible even
for solutions of the Einstein Field Equations)} \cite{HawEll73}.
\end{quote}
The EBU model solves this key problem (\S \ref{sec:chron}), which
means ordering 1. is also OK in them.\\

Unlike the Block Universe models, the General Relativity EBU models
adequately represents 4., 5., and 6., which is the same in them as
in Newtonian theory.\\

When quantum effects are significant, the future manifests all the
signs of quantum weirdness, including duality, uncertainty, and
entanglement. With the passage of time, after the time-irreversible
process of state-vector reduction has taken place, the past emerges,
with the previous quantum uncertainty replaced by the classical
certainty of definite particle identities and states. The present
time is where this transition largely takes place. But the process
does not take place uniformly or reversibly:  evidence from delayed
choice experiments shows that some isolated patches of quantum
indeterminacy remain, and their transition from probability to
certainty only takes place later. Thus, when quantum effects are
significant,  the Evolving Block Universe (``EBU") of classical
physics cedes way to the Crystallizing Block Universe (``CBU")
\cite{EllRot10}. On large enough scales that quantum effects are not
significant, the two models become indistinguishable.\\

\noindent Interesting work to be done arising of the EBU proposal,
apart from testing its basic ideas, includes
\begin{enumerate}
     \item Determining the nature of the preferred time surfaces
     defined in \S \ref{sec:surf} in inhomogeneous cosmologies (they
     are the same as the usual surfaces in spatially homogeneous
     models);
     \item Extending the ADM analysis of \S \ref{sec:rate} to the
     case where the preferred surfaces of constant time go null and then become
     timelike;
     \item Relating the geometry of time surfaces when spacetime is represented on different
     averaging scales; this is an aspect of the fitting and averaging
     problem for general relativity theory
     \cite{Ell84,EllSto87};
     \item Determining how the idea can sensibly relate to entanglement and EPR
     type experiments \cite{AhaRoh05,GreZaj06};
     \item Investigating the relation of quantum gravity theories to the EBU proposal.
      \end{enumerate}
      We do not yet have a reliable theory of quantum gravity, but there are
      some proposals that do indeed see time at the quantum level as unfolding in a way
      analogous to the EBU (for example spin foam models \cite{Bae98}). My view would be that
      however they relate to time \cite{Ish92,Ish93,IshBut99,IshBut00,HugVisWut12,And12},
      they must be capable of producing an EBU at the classical
      level, or they will fail the fundamental test of relating convincingly
      to the physics of ordinary everyday life.
      This is a correspondence principle for these theories.\\

\noindent \textbf{Conclusion}: I have reviewed the many arguments
against the flow of time, in particular those based in the
Wheeler-de Witt equation, and have argued that they do not carry the
day: the EBU is a good model of spacetime that fits well with our
daily experience as well as with general relativity and quantum
theory. A key issue is how the properties of time relate to the
experiences we have through the operations of our mind; I have
argued (\S \ref{sec:mind}) that this is crucial evidence we must
take into account:
\begin{quote}
\textbf{A key test}: \emph{The experimental evidence supporting the
huge corpus of present-day neuroscience \cite{KanSchJes00}
decisively favors the EBU over the usual Block Universe proposal, at
the classical level; therefore to be acceptable, any proposed
underlying theory must pass the critical test of leading to an
effective EBU at the macro level.}
\end{quote}
 The physics equations we should believe are those that are compatible
with this evidence; those that are not fail a basic reality test.\\

\textbf{Acknowledgement}: I thank R Goswami,  C Clarkson, R Tavakol, and T Clifton for
helpful comments.

\section*{Appendix: R Goswami}

This Appendix by R Goswami develops further the relation of the
proposal made here to the usual ADM formalism (see \S \ref{sec:rate}).\\

 Let us consider a globally hyperbolic
manifold (${\cal M},g$), having a topological structure $\Sigma
\otimes \textbf{R}$, where $\Sigma_t$ denotes the family of
spacelike hypersurfaces labelled by the parameter $t$. On each
hypersurface of constant $t$, we can define a purely spatial metric
as $h_{ab}=g_{ab}+n_an_b$ where $n^a$ is the (necessarily timelike)
unit normal vector of $\Sigma_t$ with $n^an_a=-1$ . Hence, given the
foliation on (${\cal M},g$), the spatial metric $h_{ab}$ on the
spacelike hypersurface $\Sigma_t$ is uniquely defined.

Let us also assume, that the Ricci tensor $R_{ab}$ on this manifold,
has one timelike and three spacelike eigenvectors. These
eigenvectors are unique for any physically realistic ({\it Type I})
non-zero matter field. Let the time-line for a given observer, be
the integral curve of the timelike eigenvector $t^a$ of $R_{ab}$.
This then uniquely defines the {\it shift}-vector with respect to a
given foliation of a family of spacelike hypersurfaces $\Sigma_t$ as
\begin{equation}
N^a=h^a_bt^b
\end{equation}
Furthermore, if we specify the relation between this co-ordinate
time `$t$' and proper time `$\tau$` as $d\tau=N(t,x^i )dt$, (where
$x^i$ are the co-ordinates on the 3-surface $\Sigma_t$), then by
definition this gives the lapse function
\begin{equation}
N(t,x^i )=-t^an_a
\end{equation}
Specifically if the co-ordinate time is equal to the proper time the
we must have $t^an_a=-1$.\\

\noindent Author emails:\\

George Ellis: gfrellis@gmail.com

Rituparno Goswami: vitasta9@gmail.com

\vspace{0.25in}
 \small{Version 2012-08-26}.


\begin{thebibliography}{99}

\bibitem{AhaRoh05} Y Aharanov and D Rohrlich (2005): \emph{Quantum Paradoxes. Quantum Theory
for the Perplexed} (Weinheim: Wiley-VCH Verlag).

\bibitem{And12} Edward Anderson (2012) ``Problem of Time in
Quantum Gravity'': arXiv:1206.2403.

\bibitem{And72} P W Anderson (1972) ``More is different''
\emph{Science} \textbf{177}: 393-396.

\bibitem{Ann01} Peter Anninos (2001)
``Computational Cosmology: From the Early Universe to the Large
Scale Structure'' \emph{Living Rev. Relativity} \textbf{4}:2
[http://relativity.livingreviews.org/Articles/lrr-2001-2/].

\bibitem{ADM} Richard Arnowitt, Stanley Deser and Charles W. Misner (1962) ``The
Dynamics of General Relativity''. In \emph{Gravitation: An
Introduction to Current Research}, Ed. Louis Witten (Wiley):
227-265. Reprinted in \emph{Gen. Rel. Grav}. \textbf{40}: 1997
(2008).

\bibitem{Bae98} John C. Baez (1998) ``Spin Foam Models''
\emph{Class.Quant.Grav}. \textbf{15} (1998) 1827-1858 [arXiv:gr-qc/9709052v3].

\bibitem{Bar99} J B Barbour (1999) \emph{The End of Time: The Next Revolution
in Phyiscs} (Oxford: Oxford University Press).

\bibitem{BrePet06} H.-P Breuer and F Petruccione (2006) \emph{The Theory of open quantum
systems} (Oxford: Clarendon Press).

\bibitem{BriRos00} J.S. Briggs and J.M. Rost: ``Time dependence in quantum mechanics''
 \emph{Eur. Phys. J}. \textbf{D 10}, 311-318.

\bibitem{BriRos01} John S. Briggs  and Jan M. Rost (2001) ``On the Derivation of the
Time-Dependent Equation of Schro\"{o}dinger'' \emph{Foundations of
Physics}: \textbf{31}:693-712.

\bibitem{Bro23} C D Broad (1923), \emph{Scientific Thought} (New York: Harcourt, Brace and Co.).
For Table of Contents and some chapters, see
http://www.ditext.com/broad/st/st-con.html.

\bibitem{But10}
J. N. Butterfield (2010) ``The End of Time?'' [arXiv:gr-qc/0103055].

\bibitem{IshBut99}
 J. Butterfield and C.J.Isham (1999) ``On the Emergence of Time in
Quantum Gravity'' In \emph{The Arguments of Time}, ed. J.
Butterfield, Oxford University Press, 1999 [arXiv:gr-qc/9901024v1].

\bibitem{IshBut00}
J.Butterfield and C.J.Isham (2000) ``Spacetime and the Philosophical
Challenge of Quantum Gravity'' In \emph{Physics meets Philosophy at
the Planck Scale}, ed. C. Callender and N. Huggett, Cambridge
University Press (2000) [arXiv:gr-qc/9903072v1].

\bibitem{Car10} S Carroll (2010) \emph{From Eternity to here: the quest
for the ultimate arrow of time} (New York: Dutton).

\bibitem{Dav74} P C W Davies (1974), \emph{The physics of time asymmetry}. (Surrey University Press, London).

\bibitem{Dav12} P C W Davies (2012), ``That Mysterious Flow''. \emph{Scientific
American} Special Edition: \emph{A Matter of Time} Vol \textbf{21}:
2012), 8-13.

\bibitem{Dod03} S Dodelson (2003) \emph{Modern Cosmology} (New York: Academic Press).

\bibitem{Edd28} A S Eddington (1928). \emph{The Nature of the Physical World}.
(London: MacMillan).

\bibitem{Ell67} G F R Ellis (1967)  ``The dynamics of pressure-free matter in
general relativity".  \emph{Journ Math Phys} \textbf{8}, 1171-1194.

\bibitem{Ell71} G F R Ellis (1971) ``Relativistic Cosmology". In
\emph{General Relativity and
Cosmology}, Proc Int School of Physics ``Enrico Fermi" (Varenna),
Course XLVII. Ed. R K Sachs (Academic Press, 1971), 104-179.
Reprinted as Golden Oldie:: \emph{Gen. Rel. Grav}. \textbf{41}: 581
(2009).

\bibitem{Ell84} G F R Ellis (1984): ``Relativistic cosmology: its nature, aims and
problems".In \emph{General Relativity and Gravitation}, Ed B
Bertotti et al (Reidel, 215-288.

\bibitem{Ell02}  G F R Ellis (2002) ``Cosmology and Local Physics''. \emph{New Astronomy Reviews}
\textbf{46}: 645-658  ( gr-qc/0102017).

\bibitem{Ell06} G F R Ellis (2006) ``Physics in the Real
Universe: Time and Spacetime''. \emph{GRG } \textbf{38}:1797-1824
[arXiv:gr-qc/0605049].

\bibitem{Ell08} G  F R Ellis (2008) ``On the nature of causation in
complex systems'' \emph{Trans Roy Soc South Africa} \textbf{63}:
69-84 .

\bibitem{Ell11a} G  F R Ellis (2012) ``On the limits of quantum theory:
contextuality and the quantum-classical cut'' \emph{Annals of
Physics} \textbf{327} 1890-1932 [arXiv:1108.5261].

\bibitem{Ell11}
George F R Ellis (2011) ``The arrow of time, the nature of
spacetime, and quantum measurement''
\begin{verbatim}
 http://www.mth.uct.ac.za/~ellis/Quantum_arrowoftime_gfre.pdf
\end{verbatim}

\bibitem{KinEll73} A R King and G F R Ellis (1973): ``Tilted homogeneous cosmologies".
\emph{Comm Math Phys} \textbf{31}, 209-242.

\bibitem{EllMat85} G F R Ellis and D R Matravers: ``Spatial Homogeneity and the size of
the universe". In \emph{A Random Walk in Relativity and Cosmology}
(Raychaudhuri Festschrift). Ed N Dadhich, J K Rao, J V Narlikar, and
C V Vishveshswara. (Wiley Eastern, Delhi. 1985), 92-108.

\bibitem{EllRot10} G F R Ellis and T Rothman (2010): "Crystallizing block universes".
\emph{International Journal of Theoretical Physics} \textbf{49}:
988. [arXiv:0912.0808].

\bibitem{EllSci72} G F R Ellis and D W Sciama: ``Global and non-global problems in
cosmology". In \emph{General Relativity (A Synge Festschrift)}, ed.
L. O'Raifeartaigh (Oxford University Press, 1972), 35-59.

\bibitem{EllSto87} G F R Ellis and W R Stoeger  (1987): ``The Fitting Problem in
Cosmology". \emph{Class Qu Grav} \textbf{4}, 1679-1690.

\bibitem{Elletal11} George F. R. Ellis, Henk van Elst, Jeff Murugan, Jean-Philippe Uzan
(2011) "On the Trace-Free Einstein Equations as a Viable Alternative
to General Relativity"  \emph{Class. Quantum Grav}. \textbf{28}:
225007 [arXiv:1008.1196].

\bibitem{EllWil00} G F R Ellis and R M Williams (2000) \emph{Flat and Curved Space
Times}. (Oxford University Press, Second edition).

\bibitem{Fey85} R Feynman (1985) \emph{QED: The Strange Theory of Light and Matter}
(Princeton: Princeton University Press).

\bibitem{FeyLeiSan63} R P Feynman, R B Leighton and M Sands (1963)
\emph{The Feynman lecturs on Physics: Mainly Mechanics, Radiation,
and Heat} (Reading, Mass: Addison-Wesley).

\bibitem{FeyLeiSan64} R P Feynman, R B Leighton and M Sands (1964)
\emph{The Feynman lecturs on Physics: The Electromagnetic Field}
(Reading, Mass: Addison-Wesley).

\bibitem{FeyLeiSan65} R P Feynman, R
B Leighton and M Sands (1965) \emph{The Feynman lecturs on Physics:
Quantum Mechanics} (Reading, Mass: Addison-Wesley).

\bibitem{fqxi}
FQXI essay competition (2010):
http://fqxi.org/community/forum/category/10.

\bibitem{fqxi_meet}
FQXI meeting on time (2011): http://fqxi.org/conference/2011.

\bibitem{GamPul12} R Gambini and J Pullin (2012) \emph{A First Course in Loop Quantum Gravity} (Oxford: Oxford University Press).

\bibitem{GemMicMah04} J Gemmer, M Michel and G Mahler (2004) \emph{Quantum
Thermodynamics: Emergence of Thermodynamic Behaviour Within
Composite Quantum Systems} (Heidelberg: Springer).

\bibitem{GibSheRan03} G W Gibbons, E P S Shellard and S J Rankin (Eds) (2003) \emph{The
Future of Theoretical physics and Cosmology: Celebrating Stephen
Hawking's 60th Birthday} (Cambridge: Cambridge University Press).

\bibitem{GreZaj06} G Greenstein and A G Zajonc (2006) \emph{The Quantum Challenge: Modern
Research on the Foundations of Quantum Mechanics} (Sudbury, Mass:
Jones and Bartlett).

\bibitem{Hal03} J Halliwell (2003) ``The interpretation of quantum cosmology and the problem of time''. In \emph{The
Future of Theoretical phyiscs and Cosmology: Celebrating Stephen
Hawking's 60th Birthday}, Ed. G W Gibbons, E P S Shellard and S J
Rankin (Cambridge: Cambridge University Press), 675-690.

\bibitem{Har03} J Hartle (2003) ``Theories of everything and Hawking's wave function''.
In \emph{The
Future of Theoretical phyiscs and Cosmology: Celebrating Stephen
Hawking's 60th Birthday}, Ed. G W Gibbons, E P S Shellard and S J
Rankin (Cambridge: Cambridge University Press), 38-49 and 615-620.

\bibitem{Haw84} S W Hawking (1984) ``The quantum state of the universe''
\emph{Nucl. Phys}. \textbf{B239}, 2447.

\bibitem{Haw92} S W Hawking (1992) ``The chronology protection conjecture''.
\emph{Phys. Rev}. \textbf{D46}, 603-611.

\bibitem{HawEll73} S W Hawking and G F R Ellis (1973):
\emph{The Large Scale Structure of Space-Time}. (Cambridge:
Cambridge University Press).

\bibitem{HugVisWut12} Nick Huggett, Tiziana Vistarini, and Christian Wuthrich (2012)
``Time in quantum gravity''. To appear,  \emph{The Blackwell
Companion to the Philosophy of Time}, Ed Adrian Bardon and Heather
Dyke [arXiv:1207.1635].

\bibitem{Ish92}
Chris J. Isham (1992) ``Canonical Quantum Gravity and the Problem of
Time'' Lectures at the NATO Summer School held in Salamanca
[gr-qc/9210011].

\bibitem{Ish93}
C.J.Isham (1993) ``Prima Facie Questions in Quantum Gravity''
 arXiv:gr-qc/9310031v1

\bibitem{Ish97} C J Isham (1997) \emph{Lectures on Quantum Theory: Mathematical and Structural
Foundations} (London: Imperial College Press).

\bibitem{Jacetal07} Vincent Jacques, E. Wu, Frederic Grosshans, Francois Treussart,
Philippe Grangier, Alain Aspect, Jean-Francois Roch  (2007).
``Experimental realization of Wheeler's delayed-choice
GedankenExperiment'' \emph{Science} \textbf{315}, 5814
[arXiv:quant-ph/0610241v1].

\bibitem{KanSchJes00} E R Kandel, J H Schwartz, and T M Jessell (2000)
\emph{Principles of Neuroscience} (New York: McGraw Hill).

\bibitem{rand} I Kanter, Y Aviad, I Reidler, E Cohen, and M Rosenbluh (2010). ``An optical ultrafast random bit generator''.
\emph{Nature Photonics}, \textbf{4}: 58–61.

\bibitem{Kimetal00} Yoon-Ho Kim, R. Yu, S.P. Kulik, Y.H. Shih, Marlan O. Scully [2000]
``A Delayed Choice Quantum Eraser'' \emph{Phys.Rev.Lett}.
\textbf{84}:1-5 [arXiv:quant-ph/9903047v1].

\bibitem{KupNikZla12} Oleg Kupervasser, Hrvoje Nikoli, Vinko Zlati (2012) ``The Universal Arrow of Time'': arXiv:1011.4173.

\bibitem{Mel98} D H Mellor, \emph{Real Time II}. (Routledge, London:
1998).

\bibitem{Mot31} N.F. Mott (1931) ``Time dependence in quantum mechanics''
\emph{Proc. Camb. Phil. Soc}. \textbf{27}, 553.

\bibitem{Nar} J V Narlikar (1996) \emph{The lighter side of gravity} (Cambridge University Press).

\bibitem{Pen89} R Penrose (1989) \emph{The Emperor's New Mind} (Oxford: Oxford University Press).

\bibitem{Pen04} R Penrose (2004) \emph{The Road to Reality: A complete guide to the
Laws of the Universe} (London: Jonathan Cape).

\bibitem{Pen11} R Penrose (2011) \emph{Cycles of Time: An Extraordinary New View of the Universe}
(New York: Knopf).

\bibitem{Per91} I Percival  (1991), ``Schr\"{o}dinger's quantum cat''. \emph{Nature} \textbf{351},
357.

\bibitem{Per97} Asher Peres (1997) ``Critique of the Wheeler-DeWitt equation'' In
\emph{On Einstein's Path} ed. by A. Harvey (Springer, 1998) pp.
367-379 [arXiv:gr-qc/9704061v2].

\bibitem{Pri96} H Price (1996) \emph{Time's Arrow and Archimedes' Point} (New York: Oxford
University Press).

\bibitem{Ree95} M J Rees (1995) \emph{Perspectives in astrophysical
cosmology} (Cambridge: Cambridge University Press).

\bibitem{Rov08} C Rovelli (2008) "Forget time" FQXI essay,
http://fqxi.org/community/forum/topic/237.

\bibitem{Sav01} S Savitt, ``Being and Becoming in Modern Physics'', In
\emph{The Stanford Encyclopedia of Philosophy} (Spring 2002
Edition), Edward N. Zalta (ed.), see
http://plato.stanford.edu/archives/spr2002/entries/spacetime-bebecome/.

\bibitem{ScaWheWil01} J Scalo, J Craig Wheeler and P Williams (2001),
�Intermittent jolts of galactic UV radiation: Mutagenetic effects�
In \emph{Frontiers of Life; 12th Rencontres de Blois}, ed. L. M.
Celnikier [astro-ph/0104209].

\bibitem{SciAm12} \emph{Scientific American} Special Edition (2012):
\emph{A Matter of Time} Vol \textbf{21}: 8-13.

\bibitem{Smo09} Lee Smolin (2009) ``Quantization
of unimodular gravity and the cosmological constant problems''
\emph{Phys.\ Rev.} D \textbf{80} 084003 [arXiv:0904.4841v1 [hep-th].

\bibitem{Smo10}
 Lee Smolin (2010)`` Unimodular loop quantum gravity and the problems of time'':
 arXiv:1008.1759.

\bibitem{Sob95} Dava Sobel (1995) \emph{Longitude: The True Story of a Lone Genius
Who Solved the Greatest Scientific Problem of His Time} (Walker and
Company).

\bibitem{Sor94} R.D. Sorkin,
`On the Role of Time in the Sum-over-histories Framework for Gravity
Int.'' J. Theor. Phys. 33:523-534 (1994); R Sorkin Spacetime and
causal sets (1991)
http://www.cdms.syr.edu/~sorkin/some.papers/66.cocoyoc.pdf.

\bibitem{Spi67} Murray R. Spiegel (1967) \emph{Theory and Problems of Theoretical
Mechanics} (Schaum - McGraw-Hill).

\bibitem{Tho95}
K S Thorne )(1995) \emph{Black Holes and Time Warps: Einstein's
Outrageous Legacy} (W. W. Norton and Company).

\bibitem{Vis02}
M Visser (2002) ``The quantum physics of chronology protection''. In
\emph{The Future of Theoretical Physics and Cosmology: Celebrating
Stephen Hawking's 60th Birthday} Ed G W Gibbons, E P S Shellard and
S J Rankin (Cambridge: Cambridge University Press), 161-173
[arXiv:gr-qc/0204022v2].

\bibitem{Unr89}
W. G. Unruh (1989) ``A Unimodular Theory Of Canonical Quantum Gravity''
\emph{Phys.Rev}.\textbf{D40}: 1048.

\bibitem{UnrWal89}
W. G. Unruh and R.M.Wald (1989) ``Time And The Interpretation Of Canonical
Quantum Gravity''. \emph{Phys.Rev.} \textbf{D40}:2598.

\bibitem{Wei89}
Steven Weinberg (1989) ``The cosmological constant problem''
\emph{Rev.\ Mod.\ Phys.} \textbf{61} 1--23. 13.

\bibitem{Whe78} John Archibald Wheeler (1978), ``The 'Past' and the 'Delayed-Choice
Double-Slit Experiment','' pp 9-48 in A.R. Marlow, editor,
\emph{Mathematical Foundations of Quantum Theory}, Academic Press.

\bibitem{WheFey45} J. A. Wheeler and R. P. Feynman (1945), ``Interaction with the
Absorber as the Mechanism of Radiation''. \emph{Rev. Mod. Phys}.
\textbf{17}, 157-181 .

\bibitem{Zeh07} H-D Zeh (2007) \emph{The Physical Basis of the Direction of Time} (Berlin: Springer
Verlag).

\end{thebibliography}
\end{document}